\pdfoutput=1

\documentclass[aps,prd,preprint,titlepage,nofootinbib,amsmath,amsfonts]{revtex4-2}
\usepackage[colorlinks,allcolors=blue]{hyperref}
\usepackage[T1]{fontenc}
\usepackage[USenglish]{babel}
\usepackage{bm}
\usepackage{bbm}
\usepackage{enumerate}
\usepackage{multirow}

\newcommand{\cg}[6]{
\biggl\langle
\begin{matrix}
#1	& #3	\\[-5pt]
#2	& #4
\end{matrix}
\bigg\vert
\begin{matrix}
#5	\\[-5pt]
#6 
\end{matrix}
\biggr\rangle
}

\newcommand{\smallcg}[6]{
\bigl\langle
\begin{smallmatrix}
#1	& #3	\\
#2	& #4	\\
\end{smallmatrix}
\big\vert
\begin{smallmatrix}
#5	\\
#6 	\\
\end{smallmatrix}
\bigr\rangle
}

\newcommand{\vket}[4]{
\biggl\vert
#1
\begin{matrix}
#2	\\[-5pt]
#3
\end{matrix}
#4
\biggr\rangle
}

\newcommand{\smallvket}[4]{
\Bigl\vert
#1
\begin{smallmatrix}
#2	\\[.2em]
#3
\end{smallmatrix}
#4
\Bigr\rangle
}

\newcommand{\vbra}[4]{
\biggl\langle
#1
\begin{matrix}
#2	\\[-5pt]
#3
\end{matrix}
#4
\biggr\vert
}

\begin{document}

\title{Model-independent predictions for decays\texorpdfstring{\\}{ }of hidden-heavy hadrons into pairs of heavy hadrons}
\author{E. Braaten}
\email{braaten.1@osu.edu}
\affiliation{Department of Physics, The Ohio State University, Columbus, OH 43210, USA}
\author{R. Bruschini}
\email{bruschini.1@osu.edu}
\affiliation{Department of Physics, The Ohio State University, Columbus, OH 43210, USA}

\begin{abstract}
Hidden-heavy hadrons can decay into pairs of heavy hadrons through transitions from confining Born-Oppenheimer potentials to hadron-pair potentials with the same Born-Oppenheimer quantum numbers. The transitions are also constrained by conservation of angular momentum and parity. From these constraints, we derive model-independent selection rules for decays of hidden-heavy hadrons into pairs of heavy hadrons. The coupling potentials are expressed as sums of products of Born-Oppenheimer transition amplitudes and angular-momentum coefficients. If there is a single dominant Born-Oppenheimer transition amplitude, it factors out of the coupling potentials between hidden-heavy hadrons in the same Born-Oppenheimer multiplet and pairs of heavy hadrons in specific heavy-quark-spin-symmetry doublets. If furthermore the kinetic energies of the heavy hadrons are much larger than their spin splittings, we obtain analytic expressions for the relative partial decay rates in terms of Wigner 6-$j$ and 9-$j$ symbols. We consider in detail the decays of quarkonia and quarkonium hybrids into the lightest heavy-meson pairs. For quarkonia, our model-independent selection rules and relative partial decay rates agree with previous results from quark-pair-creation models in simple cases and give stronger results in other cases. For quarkonium hybrids, we find disagreement even in simple cases.
\end{abstract}

\maketitle
\section{Introduction}

The field of hidden-heavy-hadron spectroscopy began in 1974 with the discovery of the 
hidden-charm meson $J\!/\!\psi$ \cite{Aub74,Aug74}. By the end of the century, dozens of hidden-charm and hidden-bottom mesons had been discovered, all of which matched the pattern of quarkonium energy levels predicted by quark models. The spectrum of hidden-heavy hadrons was generally believed to be well understood. But all this changed in 2003 with the discovery of the $X(3872)$ meson \cite{Cho03}. Its observed decays revealed that it is a hidden-charm meson whose constituents include an additional light quark-antiquark pair. Since then, dozens of exotic hidden-charm and hidden-bottom hadrons have been discovered, inaugurating a \textit{renaissance} of hidden-heavy-hadron spectroscopy; see, for instance, the reviews in Refs.~\cite{Leb17,Bra20}.

In this paper, we focus on decays of hidden-heavy hadrons into pairs of heavy hadrons. These strong decays are subject to the low energy, nonperturbative regime of QCD, which is notoriously difficult to access from first principles. Hadron resonances with 2-body (and some 3-body) strong decay channels can be calculated \textit{ab initio} using lattice QCD and the L\"uscher formalism; see Ref.~\cite{Bri17} for a review. But hidden-heavy hadrons that are above the threshold for a heavy-hadron pair have many multibody strong decay channels, including lower-energy hidden-heavy hadrons plus light hadrons. So, first-principles calculations of hidden-heavy hadron resonances using the L\"uscher formalism might be impractical.

An alternative approach to the study of hidden-heavy hadrons from first principles is the \emph{Born-Oppenheimer approximation} for QCD \cite{Jug99}. The Born-Oppenheimer (B\nobreakdash-O) approximation is based on the assumption that gluons and light quarks respond almost instantaneously to the motion of the heavy quark-antiquark pair,  since the latter have a mass $m_Q$ that is much larger than the nonperturbative energy scale of QCD $\Lambda_\textup{QCD}$. In this approximation, the energy levels of the gluon and light-quark fields in the presence of static quark and antiquark sources, which can be calculated using lattice QCD, play the role of \emph{B\nobreakdash-O potentials} that determine the motion of the heavy quark-antiquark pair through a Schr\"odinger equation. It has been shown that the B\nobreakdash-O approximation can be formulated as a rigorous effective theory of QCD \cite{Ber15,On17,Bra18a,Sot20}.

There are two qualitatively different kinds of B\nobreakdash-O potentials in QCD. The first kind consists of confining potentials that increase linearly at large distances \cite{Jug03}. The solutions to the Schr\"odinger equation in these potentials are bound states associated with compact hidden-heavy hadrons. The second kind consists of potentials that at large distances approach the scattering threshold of a pair of heavy hadrons \cite{Bal05}. The solutions to the Schr\"odinger equation in these potentials include scattering states of heavy-hadron pairs. They might also include bound states associated with heavy-hadron molecules.

The B\nobreakdash-O potentials for hidden-heavy hadrons and for heavy-hadron pairs mix with each other \cite{Bal05,Bul19}. Hence, any hidden-heavy hadron is associated with a pole of the S-matrix for heavy-hadron pairs. For hidden-heavy hadrons that are stable against decays into heavy-hadron pairs, the imaginary part of the pole is proportional to the total decay width into lower-energy hidden-heavy hadrons plus light hadrons. For hidden-heavy hadrons associated with heavy-hadron-pair scattering resonances, the imaginary part of the pole has additional contributions from decays into heavy-hadron pairs.

One way to calculate partial decay widths of hidden-heavy hadrons into heavy-hadron pairs using the B\nobreakdash-O approximation for QCD is to calculate self-energy corrections to ``bare'' hidden-heavy-hadron states by resummation of heavy-hadron loop diagrams, as was done in Refs.~\cite{Bru21a,Tar22}. But it should be noted that this approach can lead to inconsistent results for hidden-heavy hadrons that are strongly coupled to a nearby heavy-hadron-pair threshold \cite{Bru21a}. Another way is to calculate resonances directly from the Schr\"odinger equation with coupled hidden-heavy-hadron and heavy-hadron-pair channels, as was done in Refs.~\cite{Bic20,Bru21c,Leb23}. This approach should be preferred as it respects unitarity and it yields consistent results for resonances that are close to a heavy-hadron-pair threshold \cite{Bru21c}.

In general, calculating partial decay widths in the B\nobreakdash-O approximation requires solving a Schr\"odinger equation where the potentials and the transition amplitudes between them are determined using inputs from lattice QCD and/or models \cite{Bic20,Bru20,Bru21a,Bru21c,Tar22,Leb23}. However, it is possible to derive model-independent results using only B\nobreakdash-O symmetries and angular-momentum algebra, which is the main objective of this study.

The remainder of this paper is organized as follows. In Section~\ref{staticheavysec}, we define static hadrons and heavy hadrons. We then construct heavy-hadron-pair states and discuss their symmetries. In Section~\ref{decaysec}, we derive general expressions for the coupling potentials between hidden-heavy hadrons and heavy-hadron pairs as sums of products of B\nobreakdash-O transition amplitudes and angular-momentum coefficients. From these expressions, we derive model-independent selection rules for decays of hidden-heavy hadrons. In some cases, we also obtain analytic predictions for relative partial decay rates. In Section~\ref{applicationsec}, we discuss in detail the decays of quarkonia and quarkonium hybrids into pairs of heavy mesons. In Section~\ref{modelsec}, we compare our results with previous ones from quark-pair-creation models. Finally, we summarize these results in Section~\ref{conclusionsec}.

\section{Static and Heavy Hadrons}
\label{staticheavysec}

In this section, we introduce notation for the angular-momentum/parity states of a heavy hadron, a pair of static hadrons, and a pair of heavy hadrons. We also describe the symmetries of these states.

\subsection{Single Heavy Hadron}

QCD is a quantum field theory with gluon fields and quark fields. The symmetries of QCD include rotations generated by the total angular momentum vector $\bm{J}$, parity $\mathcal{P}$, and charge conjugation $\mathcal{C}$. The quark fields relevant to hadrons are those for the three light flavors  $u$, $d$, and $s$ and  the two heavy flavors $c$ and $b$. We refer to the field theory with only the gluon fields and the light quark fields as \emph{light QCD}. We denote the angular-momentum vector that generates rotations of the light-QCD fields by $\bm{J}_\textup{light}$.

A \emph{heavy hadron} is one that contains a single heavy quark ($Q$) or antiquark ($\bar{Q}$). The heavy hadron also has light-QCD fields that combine with the $Q$ or $\bar{Q}$ to produce a color singlet. A heavy hadron that contains $Q$ (or $\bar{Q}$) can be a \emph{heavy meson}, with the light-QCD fields having the flavor of a single light antiquark $\bar{q}$ (or a single light quark $q$). It can also be a \emph{heavy baryon} (or a \emph{heavy antibaryon}), with the light-QCD fields having the flavor of  two light quarks $q_1q_2$ (or  two light antiquarks $\bar{q}_1\bar{q}_2$).

A heavy quark $Q$ (or antiquark $\bar{Q}$) has spin $\frac{1}{2}$ and intrinsic parity even (or odd). Its spin/parity states are denoted by a ket $\bigl\lvert \frac{1}{2}^\pm, m\bigr\rangle$ in the text and by a ket $\smallvket{}{\frac{1}{2}^\pm}{m}{}$ in equations, where $m = +\frac{1}{2},-\frac{1}{2}$ and the two choices for the parity superscript $\pm$ correspond to $Q$ or $\bar{Q}$. The doublet corresponding to the 2 values of $m$ will be labeled more concisely as $\frac{1}{2}^\pm$. The transformations under $\mathcal{P}$ and $\mathcal{C}$ of the spin/parity state $\frac{1}{2}^\pm$ of a heavy quark or antiquark are:
\begin{subequations}
\label{cpheavy}
\begin{align}
\mathcal{P} \vket{}{\frac{1}{2}^\pm}{m}{} &= \pm  \vket{}{\frac{1}{2}^\pm}{m}{}, \label{pheavy} \\
\mathcal{C} \vket{}{\frac{1}{2}^\pm}{m}{} &=  \vket{}{\frac{1}{2}^\mp}{m}{}. \label{cheavy}
\end{align}
\end{subequations}

In the heavy-quark limit, the spin of the heavy quark $Q$ (or antiquark $\bar{Q}$) decouples from the light-QCD fields. As far as the light-QCD fields are concerned, the $Q$ (or  $\bar{Q}$) reduces to a static color-triplet ($\bm{3}$) (or color-antitriplet ($\bm{3}^\ast$)) source. The light-QCD fields bound to the static color source  form a color singlet. We refer to the light-QCD fields together with the static color source as a \emph{static hadron}. If the light-QCD fields have flavor $\bar{q}$ (or $q$), it is a \emph{static meson}. If the light-QCD fields have flavor $q_1 q_2$ (or $\bar{q}_1\bar{q}_2$), it is a \emph{static baryon} (or a \emph{static antibaryon}).

A static hadron can be labeled by its light-QCD angular-momentum quantum numbers $(j, m)$ and by its parity eigenvalue $\pi$. We denote the angular-momentum/parity state of the static hadron by a ket $\lvert j^{\pi}, m \rangle$ in the text and by a ket $\smallvket{}{j^{\pi}}{m}{}$ in equations. The multiplet corresponding to the $2j+1$ values of $m$ will be labeled more concisely as  $j^{\pi}$. The transformations under $\mathcal{P}$ and $\mathcal{C}$ of the angular-momentum/parity state $j^{\pi}$ of a static hadron are:
\begin{subequations}
\label{cplight}
\begin{align}
\mathcal{P} \vket{}{j^{\pi}}{m}{} &= \pi  \vket{}{j^{\pi}}{m}{}, \label{plight} \\
\mathcal{C} \vket{}{j^{\pi}}{m}{} &=  \vket{}{j^{\pi (-1)^{2j}}}{m}{}. \label{clight}
\end{align}
\end{subequations}
Note that a static hadron and its charge conjugate have the same (opposite) parity if $j$ is an integer (half-integer), hence the multiplicative factor $(-1)^{2j}$ in the parity superscript on the right side of Eq.~\eqref{clight}.

A heavy-hadron state can be obtained by taking the direct product of the spin/parity state $\frac{1}{2}^\pm$ of a heavy quark or antiquark and the angular-momentum/parity state $j^{\pi}$ of a static hadron. The total angular momentum vector is the sum of $\bm{J}_\textup{light}$ and the spin of the heavy quark or antiquark. The heavy-hadron state with spin $J$ is
\begin{equation}
\vket{\Bigl(\tfrac{1}{2}^{\pm}, j^{\pi} \Bigr)}{J}{M}{} = \sum_{m, m^\prime} \cg{\tfrac{1}{2}}{m}{j}{m^\prime}{J}{M} \vket{}{\tfrac{1}{2}^\pm}{m}{} \vket{}{j^{\pi}}{m^\prime}{},
\label{hadef}
\end{equation}
where $j$ can be $J+\frac{1}{2}$ or $\bigl\lvert J - \frac{1}{2}\bigr\rvert$ and $\smallcg{j_1}{m_1}{j_2}{m_2}{j_3}{m_3}$ is a Clebsch-Gordan coefficient. The hadron is a meson if $J$ is an integer and a baryon if $J$ is a half-integer. We have adopted the standard order of writing the spin/parity of the heavy quark or antiquark first and the angular-momentum/parity of the light-QCD fields second. From Eqs.~\eqref{cpheavy}-\eqref{cplight}, the transformations under $\mathcal{P}$ and $\mathcal{C}$ of the heavy-hadron state in Eq.~\eqref{hadef} are:
\begin{subequations}
\label{cphadron}
\begin{align}
\mathcal{P} \vket{\Bigl(\tfrac{1}{2}^\pm, j^{\pi}  \Bigr)}{J}{M}{} &= \pm \pi \vket{\Bigl(\tfrac{1}{2}^\pm, j^{\pi}  \Bigr)}{J}{M}{}, \\
\mathcal{C} \vket{\Bigl(\tfrac{1}{2}^\pm, j^{\pi}  \Bigr)}{J}{M}{} &= \vket{\Bigl(\tfrac{1}{2}^\mp, j^{\pi(-1)^{2j}}\Bigr)}{J}{M}{}.
\end{align}
\end{subequations}

The parity of the heavy hadron is the product of the parity of $Q$ or $\bar Q$ and the parity of the static hadron:  $P=\pm \pi$. Note that the charge conjugate of the heavy hadron has parity $\mp\pi(-1)^{2j}=P (-1)^{2 J}$, where we have used the fact that $(-1)^{2 j} = - (-1)^{2 J}$. Since $J$ is an integer for a meson, its charge conjugate has the same parity. Since $J$ is a half-integer for a baryon, its charge conjugate has opposite parity.

Since the spin of a heavy quark or antiquark is conserved up to corrections suppressed by $1/m_Q$, the mass difference between two heavy hadrons corresponding to the same static hadron $j^{\pi}$ but with different spins $J=\bigl\lvert j - \frac{1}{2}\bigr\rvert$ and $J=j+\frac{1}{2}$ must also be suppressed by $1/m_Q$. Therefore, heavy hadrons form approximately degenerate doublets of \emph{heavy quark spin symmetry} (HQSS) labeled by the light-QCD angular-momentum/parity $j^\pi$ and other quantum numbers.

\subsection{Pair of Static Hadrons}
\label{twostaticsec}

Let us now consider a pair of static hadrons at the positions $+\frac{1}{2}\bm{r}$ and $-\frac{1}{2}\bm{r}$ with light-QCD quantum numbers $j_1^{\pi_1}$ and $j_2^{\pi_2}$, respectively. We take the first static hadron to contain the $\bm{3}$ source and the second to contain the $\bm{3}^\ast$ source.  Since the angular-momentum/parity quantum numbers correspond to rotations and reflections around different points in space, we specify the position $\pm\frac{1}{2}\bm{r}$ by an argument $(\pm)$.  The static-hadron-pair states are therefore labeled by $\bigl(j_1^{\pi_1}(+), j_2^{\pi_2}(-)\bigr)$.

Light QCD in the presence of static $\bm{3}$ and $\bm{3}^\ast$ sources at $+\frac{1}{2}\bm{r}$ and $-\frac{1}{2}\bm{r}$ has cylindrical symmetries consisting of rotations around the $\bm{\hat{r}}$ axis and reflections through planes containing $\bm{\hat{r}}$. We denote the reflection corresponding to a specific plane (which need not be specified) by $\mathcal{R}$.  It also has a discrete symmetry under $\mathcal{C}\mathcal{P}$, the combination of charge-conjugation and parity. We refer to the group formed by the cylindrical symmetries and the $\mathcal{C}\mathcal{P}$ symmetry as the B\nobreakdash-O symmetry group.

The B\nobreakdash-O symmetries imply that the eigenstates of the light-QCD Hamiltonian can be chosen to be simultaneously eigenstates of $\bm{J}_\textup{light} \cdot \bm{\hat{r}}$, that is, the projection of the light-QCD angular momentum onto the axis passing through the sources. We denote its eigenvalues by $\lambda$. They can be also chosen to be simultaneously eigenstates of $\mathcal{C}\mathcal{P}$. We denote its eigenvalues by  $\eta$.  In the sector with $\lambda=0$, the reflection operator $\mathcal{R}$ can also be diagonalized. We denote its eigenvalues by $\epsilon$. 
We refer to  $\lambda$ and $\eta$ and also $\epsilon$ if $\lambda=0$ as B\nobreakdash-O quantum numbers. Alternatively, the light-QCD Hamiltonian, $\lvert\bm{J}_\textup{light} \cdot \hat{\bm{r}}\rvert$, $\mathcal{C}\mathcal{P}$, and $\mathcal{R}$ can be simultaneously diagonalized. Thus an alternative choice for the B\nobreakdash-O quantum numbers is $\lvert\lambda\rvert$, $\eta$, and $\epsilon$. It is customary to denote these B\nobreakdash-O quantum numbers using the notation $\Lambda_\eta^\epsilon$, where $\Lambda=\lvert\lambda\rvert$, the subscript $\eta$ is $g$ or $u$ if the $\mathcal{C}\mathcal{P}$  eigenvalue $\eta$ is $+1$ or $-1$, and the superscript $\epsilon$ is $+$ or $-$ if the reflection eigenvalue $\epsilon$ is  $+1$ or $-1$. If $\Lambda > 0$, the superscript $\epsilon$ is often omitted because cylindrical symmetry requires the $\Lambda_\eta^+$ and $\Lambda_\eta^-$ states to be degenerate in energy. It is commonplace to specify integer values of $\Lambda$ with an uppercase Greek letter instead of a number, according to the code $\Lambda\to\Sigma,\Pi,\Delta$ for $\Lambda=0,1,2$ and so on.

Static-hadron-pair states can be obtained by taking direct products of the static-hadron states $j_1^{\pi_1}(+)$ and $j_2^{\pi_2}(-)$.  The direct product can be decomposed into states with light-QCD angular-momentum quantum numbers $(j^\prime, \lambda)$:
\begin{equation}
\vket{\Bigl(j_1^{\pi_1} (+), j_2^{\pi_2} (-) \Bigr)}{j^\prime}{\lambda}{} = \sum_{m_1, m_2} \cg{j_1}{m_1}{j_2}{m_2}{j^\prime}{\lambda} \vket{}{j_1^{\pi_1}}{m_1}{(+)} \vket{}{j_2^{\pi_2}}{m_2}{(-)}.
\label{lightdef}
\end{equation}
We have adopted the standard order of writing the $(+)$ state first and the $(-)$ state second. Reversal of this order produces a sign:
\begin{equation}
\vket{\Bigl(j_2^{\pi_2} (-), j_1^{\pi_1} (+) \Bigr)}{j^\prime}{\lambda}{} = (-1)^{4 j_1 j_2 + j_1 + j_2 - j^\prime}\vket{\Bigl(j_1^{\pi_1} (+), j_2^{\pi_2} (-) \Bigr)}{j^\prime}{\lambda}{}.
\label{reorder+-}
\end{equation}
The sign is the product of a factor $(-1)^{j_1 + j_2 - j^\prime}$ from the symmetries of Clebsch-Gordan coefficients and a factor $(-1)^{4 j_1 j_2}$ from changing the order of the light-QCD operators.\footnote{The factor $(-1)^{4 j_1 j_2}$ is $-1$ if both static hadrons are static mesons and $+1$ otherwise.} From Eqs.~\eqref{cplight}, the transformations under $\mathcal{P}$ and $\mathcal{C}$ of the static-hadron-pair state in Eq.~\eqref{lightdef} are:
\begin{subequations}
\begin{align}
\mathcal{P} \vket{\Bigl(j_1^{\pi_1} (+), j_2^{\pi_2} (-) \Bigr)}{j^\prime}{\lambda}{} =& \pi_1 \pi_2 \vket{\Bigl(j_1^{\pi_1} (-), j_2^{\pi_2} (+) \Bigr)}{j^\prime}{\lambda}{}, \label{peq} \\
\mathcal{C} \vket{\Bigl(j_1^{\pi_1} (+), j_2^{\pi_2} (-) \Bigr)}{j^\prime}{\lambda}{} =& \vket{\Bigl(j_1^{\pi_1 (-1)^{2 j_1}} (+), j_2^{\pi_2 (-1)^{2 j_2}} (-) \Bigr)}{j^\prime}{\lambda}{}. \label{ceq}
\end{align}
\end{subequations}
On the right side of  Eq.~\eqref{peq}, the position labels $(-)$ and $(+)$ can be put into the standard order by using  Eq.~\eqref{reorder+-}.

The static-hadron-pair state in Eq.~\eqref{lightdef} is a simultaneous eigenstate of $\mathcal{C}\mathcal{P}$ when the two static hadrons are charge conjugates, which requires $j_2^{\pi_2}=j_1^{\pi_1 (-1)^{2 j_1}}$. After reordering the $(-)$ and $(+)$ states using Eq.~\eqref{reorder+-}, the $\mathcal{C}\mathcal{P}$ eigenvalue is the product of $\pi_1^2 (-1)^{2 j_1}=(-1)^{2 j_1}$ and $(-1)^{4j_1^2 + 2 j_1 - j^\prime}$. Since $2j_1(2j_1+1)$ is even, the $\mathcal{C}\mathcal{P}$ eigenvalue reduces to
\begin{equation}
\eta = (-1)^{j^\prime + 2j_1},
\label{etalighteq}
\end{equation}
where we have used the fact that $j^\prime$ is an integer since $j_1=j_2$. Outside of this special case, the static-hadron-pair state in Eq.~\eqref{lightdef} is an equal-amplitude superposition of $\eta=+1$ and $\eta=-1$. The normalized projection onto each value of $\eta$ is
\begin{equation}
\vket{\Bigl(j_1^{\pi_1} (+), j_2^{\pi_2} (-) \Bigr)}{j^\prime}{\lambda}{,\eta} \equiv \sqrt{2} \, \Pi_{\eta} \vket{\Bigl(j_1^{\pi_1} (+), j_2^{\pi_2} (-) \Bigr)}{j^\prime}{\lambda}{},
\end{equation}
where
\begin{equation}
\Pi_\eta = \frac{\mathbbm{1} + \eta\, \mathcal{C}\mathcal{P}}{2}
\label{projeq}
\end{equation}
is the projector onto $\mathcal{C}\mathcal{P}$-parity $\eta$.

The reflection through a plane containing $\bm{\hat{r}}$ can be expressed as the product of a parity transform $\mathcal{P}$ and a rotation by the angle $\pi$ around the axis perpendicular to the reflection plane. Without loss of generality, we can identify the arbitrary axis $\bm{\hat{z}}$ with $\bm{\hat{r}}$ and consider the operator $\mathcal{R}$ for reflections through the $zx$ plane,
\begin{equation}
\mathcal{R}=e^{-i \pi J_y} \mathcal{P},
\end{equation}
with $J_y$ the generator of rotations around $\bm{\hat{y}}$. The action of $\mathcal{R}$ on the static-hadron-pair state in Eq.~\eqref{lightdef} can be obtained by applying the rotation operator $e^{-i \pi J_y}$ to both sides of Eq.~\eqref{peq}. The state on the right side of Eq.~\eqref{peq} transforms as
\begin{equation}
e^{-i \pi J_y} \vket{\Bigl(j_1^{\pi_1} (-), j_2^{\pi_2} (+) \Bigr)}{j^\prime}{\lambda}{} = (-1)^{j^\prime - \lambda} \vket{\Bigl(j_1^{\pi_1} (+), j_2^{\pi_2} (-) \Bigr)}{j^\prime}{-\lambda}{}.
\label{roteq}
\end{equation}
Therefore, the static-hadron-pair state in Eq.~\eqref{lightdef} has the following transformation under $\mathcal{R}$:
\begin{equation}
\mathcal{R} \vket{\Bigl(j_1^{\pi_1} (+), j_2^{\pi_2} (-) \Bigr)}{j^\prime}{\lambda}{} = \pi_1 \pi_2 (-1)^{j^\prime - \lambda} \vket{\Bigl(j_1^{\pi_1} (+), j_2^{\pi_2} (-) \Bigr)}{j^\prime}{-\lambda}{}.
\label{refleq}
\end{equation}

\subsection{Pair of Heavy Hadrons}
\label{twoheavysec}

Let us now consider a pair of heavy hadrons at the positions $+\frac{1}{2}\bm{r}$ and $-\frac{1}{2}\bm{r}$. We take the first heavy hadron to contain the heavy quark $Q$ and the second to contain the heavy antiquark $\bar{Q}$.

The  heavy-hadron state in Eq.~\eqref{hadef} is expressed in terms of the direct product of a spin/parity state $\frac{1}{2}^+$ for a heavy quark $Q$ (or $\frac{1}{2}^-$ for a heavy antiquark $\bar{Q}$) and an angular-momentum/parity state $j^\pi$ for a static hadron. A heavy-hadron-pair state can be expressed in terms of the direct product of two spin/parity states for $Q$ and $\bar Q$ and a static-hadron-pair state of the form in Eq.~\eqref{lightdef}. The angular-momentum operator that generates rotations of the light-QCD fields and the heavy-quark and heavy-antiquark spins is the \emph{static angular-momentum} vector
\begin{equation}
\bm{J}_\textup{static} \equiv \bm{S}_Q + \bm{J}_\textup{light},
\end{equation}
where $\bm{S}_Q$ is the sum of the two spin vectors for $Q$ and $\bar{Q}$. We denote the quantum numbers for the total-heavy-spin vector $\bm{S}_Q$ by $(S_Q,m)$. For a heavy-hadron pair, the static-angular-momentum vector coincides with the sum of the spin vectors for the two heavy hadrons, $\bm{J}_\textup{static}=\bm{J}_1 + \bm{J}_2$. Therefore, we denote the quantum numbers for the static angular momentum vector $\bm{J}_\textup{static}$ by $(S,m_S)$.

The spin/parity state corresponding to $Q$ at the position $+\frac{1}{2}\bm{r}$ is $\frac{1}{2}^+(+)$. The spin/parity state corresponding to $\bar{Q}$ at the position $-\frac{1}{2}\bm{r}$ is $\frac{1}{2}^-(-)$. Their direct product can be decomposed into states with total-heavy-spin quantum numbers $(S_Q,m)$:
\begin{equation}
\vket{\Bigl(\tfrac{1}{2}^+(+), \tfrac{1}{2}^-(-)\Bigr) }{S_Q}{m}{} = \sum_{m^\prime, m^{\prime\prime}} \cg{\tfrac{1}{2}}{m^\prime}{\tfrac{1}{2}}{m^{\prime\prime}}{S_Q}{m} \vket{}{\tfrac{1}{2}^+}{m^\prime}{(+)} \vket{}{\tfrac{1}{2}^-}{m^{\prime\prime}}{(-)}.
\label{spinstate}
\end{equation}
We have adopted the standard order of writing the $(+)$ state first and the $(-)$ state second.  Reversal of this order produces a sign:
\begin{equation}
\vket{\Bigl(\tfrac{1}{2}^-(-), \tfrac{1}{2}^+(+)\Bigr)}{S_Q}{m}{} = (-1)^{S_Q} \vket{\Bigl(\tfrac{1}{2}^+(+), \tfrac{1}{2}^-(-)\Bigr) }{S_Q}{m}{}.
\label{qq+-reorder}
\end{equation}
The sign is the product of a factor $(-1)^{S_Q - 1}$ from the symmetries of Clebsch-Gordan coefficients and a factor $-1$ from changing the order of the fermionic operators for the heavy quark and antiquark.  From Eqs.~\eqref{cpheavy}, the transformations under $\mathcal{P}$ and $\mathcal{C}$ of the total-heavy-spin state in Eq.~\eqref{spinstate} are:
\begin{subequations}
\label{cpqqeq}
\begin{align}
\mathcal{P} \vket{\Bigl(\tfrac{1}{2}^+(+), \tfrac{1}{2}^-(-)\Bigr) }{S_Q}{m}{} =& - \vket{\Bigl(\tfrac{1}{2}^+(-), \tfrac{1}{2}^-(+)\Bigr)}{S_Q}{m}{}, \label{pqqeq} \\
\mathcal{C} \vket{\Bigl(\tfrac{1}{2}^+(+), \tfrac{1}{2}^-(-)\Bigr) }{S_Q}{m}{} =& \vket{\Bigl(\tfrac{1}{2}^-(+), \tfrac{1}{2}^+(-)\Bigr) }{S_Q}{m}{}.
\end{align}
\end{subequations}
On the right side of Eq.~\eqref{pqqeq}, the position labels $(-)$ and $(+)$ can be put into the standard order by using  Eq.~\eqref{qq+-reorder}.
Thus, Eqs.~\eqref{qq+-reorder} and \eqref{cpqqeq} imply that the state in Eq.~\eqref{spinstate} is an eigenstate of $\mathcal{C}\mathcal{P}$ with eigenvalue $(-1)^{S_Q+1}$.

The direct product of the total-heavy-spin state in Eq.~\eqref{spinstate} and the static-hadron-pair state in Eq.~\eqref{lightdef} can be decomposed into states with static-angular-momentum quantum numbers $(S,m_S)$:
\begin{multline}
\vket{\Bigl[\Bigl(\tfrac{1}{2}^+(+), \tfrac{1}{2}^-(-)\Bigr) S_Q,\Bigl(j_1^{\pi_1} (+), j_2^{\pi_2} (-) \Bigr) j^\prime\Bigr]}{S}{m_S}{} \\
= \sum_{m, \lambda} \cg{S_Q}{m}{j^\prime}{\lambda}{S}{m_S} \vket{\Bigl(\tfrac{1}{2}^+(+), \tfrac{1}{2}^-(-)\Bigr)}{S_Q}{m}{} \vket{\Bigl(j_1^{\pi_1} (+), j_2^{\pi_2} (-) \Bigr)}{j^\prime}{\lambda}{}.
\label{stateSS}
\end{multline}
An alternative basis for heavy-hadron-pair states that is often more convenient  can be obtained by starting from direct products of a heavy-hadron state $\bigl(\frac{1}{2}^+, j_1^{\pi_1}\bigr) J_1(+)$ at the position $+\frac{1}{2}\bm{r}$  and  a heavy-hadron state $\bigl(\frac{1}{2}^-, j_2^{\pi_2}\bigr) J_2(-)$ at the position $-\frac{1}{2}\bm{r}$, with the heavy-hadron states defined in Eq.~\eqref{hadef}. The direct product can be decomposed into states with static-angular-momentum quantum numbers $(S,m_S)$:
\begin{multline}
\vket{\Bigl[\Bigl(\tfrac{1}{2}^+, j_1^{\pi_1}  \Bigr) J_1(+), \Bigl(\tfrac{1}{2}^-, j_2^{\pi_2}\Bigr) J_2(-) \Bigr]}{S}{m_S}{} \\
= \sum_{M_1,M_2} \cg{J_1}{M_1}{J_2}{M_2}{S}{m_S} \vket{\Bigl(\tfrac{1}{2}^+, j_1^{\pi_1}  \Bigr)}{J_1}{M_1}{(+)} \vket{\Bigl(\tfrac{1}{2}^-, j_2^{\pi_2}  \Bigr)}{J_2}{M_2}{(-)},
\label{stateHH}
\end{multline}
where we have adopted the standard order of writing the $(+)$ state first and the $(-)$ state second. Reversal of this order produces a sign:
\begin{multline}
 \vket{\Bigl[\Bigl(\tfrac{1}{2}^-, j_2^{\pi_2}\Bigr) J_2(-), \Bigl(\tfrac{1}{2}^+, j_1^{\pi_1}  \Bigr) J_1(+) \Bigr]}{S}{m_S}{} \\
 = (-1)^{4 J_1 J_2 + J_1 + J_2 - S}  \vket{\Bigl[\Bigl(\tfrac{1}{2}^+, j_1^{\pi_1}  \Bigr) J_1(+), \Bigl(\tfrac{1}{2}^-, j_2^{\pi_2}\Bigr) J_2(-) \Bigr]}{S}{m_S}{}.
\label{Reorder+-}
\end{multline}
The sign is the product of a factor $(-1)^{J_1 + J_2 - S}$ from the symmetries of Clebsch-Gordan coefficients and a factor $(-1)^{4 J_1 J_2}$ from changing the order of the operators for the heavy hadrons $(\frac{1}{2}^+,j_1^{\pi_1})J_1$ and $(\frac{1}{2}^-,j_2^{\pi_2})J_2$.\footnote{The factor $(-1)^{4 J_1 J_2}$ is $-1$ if both heavy hadrons are baryons and $+1$ otherwise.}

From Eqs.~\eqref{cphadron}, the transformation properties under $\mathcal{P}$ and $\mathcal{C}$ of the hadron-pair state in Eq.~\eqref{stateHH} are:
\begin{subequations}
\begin{align}
\mathcal{P} \vket{\Bigl[\Bigl(\tfrac{1}{2}^+, j_1^{\pi_1}  \Bigr) J_1(+), \Bigl(\tfrac{1}{2}^-, j_2^{\pi_2}&\Bigr) J_2(-) \Bigr]}{S}{m_S}{} \notag \\
&= - \pi_1 \pi_2 \vket{\Bigl[\Bigl(\tfrac{1}{2}^+, j_1^{\pi_1}  \Bigr) J_1(-), \Bigl(\tfrac{1}{2}^-, j_2^{\pi_2}\Bigr) J_2(+) \Bigr]}{S}{m_S}{},
\label{pspineq} \\
\mathcal{C} \vket{\Bigl[\Bigl(\tfrac{1}{2}^+, j_1^{\pi_1}  \Bigr) J_1(+), \Bigl(\tfrac{1}{2}^-, j_2^{\pi_2}&\Bigr) J_2(-) \Bigr]}{S}{m_S}{} \notag \\
&= \vket{\Bigl[\Bigl(\tfrac{1}{2}^-, j_1^{\pi_1(-1)^{2 j_1}}  \Bigr) J_1(+), \Bigl(\tfrac{1}{2}^+, j_2^{\pi_2(-1)^{2 j_2}}\Bigr) J_2(-) \Bigr]}{S}{m_S}{}.
\label{cspineq}
\end{align}
\end{subequations}
On the right side of Eq.~\eqref{pspineq}, the position labels $(-)$ and $(+)$ can be put into the standard order by using  Eq.~\eqref{Reorder+-}.

The hadron-pair state in Eq.~\eqref{stateHH} is a simultaneous eigenstate of $\mathcal{C} \mathcal{P}$ when the two heavy hadrons are charge conjugates, which requires $J_2=J_1$ and $j_2^{\pi_2}=j_1^{\pi_1 (-1)^{2 j_1}}$. After reordering the $(-)$ and $(+)$ states using Eq.~\eqref{Reorder+-}, the sign is the product of $-\pi_1^2 (-1)^{2 j_1}=(-1)^{2 J_1}$ and $(-1)^{4 J_1^2 + 2 J_1 - S}$. Since $2 J_1 ( 2 J_1 + 1)$ is even, the $\mathcal{C}\mathcal{P}$ eigenvalue reduces to
\begin{equation}
\eta^\prime = (-1)^{S + 2 J_1},
\end{equation}
where we have used the fact that $S$ is an integer since $J_1=J_2$.  Outside of this special case, the hadron-pair state in Eq.~\eqref{stateHH} is an equal-amplitude superposition of $\eta^\prime=+1$ and $\eta^\prime=-1$. In this case, the normalized projection onto each value of $\eta^\prime$ is
\begin{multline}
\vket{\Bigl[\Bigl(\tfrac{1}{2}^+, j_1^{\pi_1}  \Bigr) J_1(+), \Bigl(\tfrac{1}{2}^-, j_2^{\pi_2}\Bigr) J_2(-) \Bigr]}{S}{m_S}{,\eta^\prime} \\
\equiv \sqrt{2} \, \Pi_{\eta^\prime} \vket{\Bigl[\Bigl(\tfrac{1}{2}^+, j_1^{\pi_1}  \Bigr) J_1(+), \Bigl(\tfrac{1}{2}^-, j_2^{\pi_2}\Bigr) J_2(-) \Bigr]}{S}{m_S}{},
\label{mesonpaircpstate}
\end{multline}
with $\Pi_{\eta^\prime}$ the projector defined in Eq.~\eqref{projeq}.

The hadron-pair states in Eq.~\eqref{stateHH} can be expanded in terms of the hadron-pair states defined in Eq.~\eqref{stateSS}:
\begin{multline}
\vket{\Bigl[\Bigl(\tfrac{1}{2}^+, j_1^{\pi_1}  \Bigr) J_1(+), \Bigl(\tfrac{1}{2}^-, j_2^{\pi_2}\Bigr) J_2(-) \Bigr]}{S}{m_S}{}
= (-1)^{2 j_1} \sqrt{\tilde{J}_1 \tilde{J}_2} \\
\times \sum_{j^\prime, S_Q} \sqrt{\tilde{\jmath}^\prime \tilde{S}_Q}
\begin{Bmatrix}
\tfrac{1}{2}		& \tfrac{1}{2}				& S_Q	\\
j_1		& j_2	& j^\prime	\\
J_1				& J_2						& S	\\
\end{Bmatrix}
\vket{\Bigl[\Bigl(\tfrac{1}{2}^+(+), \tfrac{1}{2}^-(-)\Bigr) S_Q,\Bigl(j_1^{\pi_1} (+), j_2^{\pi_2} (-) \Bigr) j^\prime\Bigr]}{S}{m_S}{},
\label{wignereq}
\end{multline}
where $\tilde{J}=2J+1$ and $\left\{\begin{smallmatrix} j_1 & j_2 & j_3 \\ j_4 & j_5 & j_6 \\ j_7 & j_8 & j_9 \end{smallmatrix}\right\}$ is a Wigner 9-$j$ symbol. The sign $(-1)^{2 j_1}$ is due to changing the order of the light-QCD operator for the static hadron $j_1^{\pi_1}$ and the fermionic operator for the heavy antiquark state $\tfrac{1}{2}^-$.

\section{Decays of a Hidden-Heavy Hadron}
\label{decaysec}

In this section, we determine the general form of the coupling potential between a confining B\nobreakdash-O potential and a hadron-pair potential. We use it to derive selection rules for decays of a hidden-heavy hadron and analytic results for relative partial decay rates.

\subsection{Coupling Potentials}

In the B\nobreakdash-O approximation, a hidden-heavy hadron state can be expanded in terms of light-QCD states in the presence of static $\bm{3}$ and $\bm{3}^\ast$ sources at the positions $+\frac{1}{2}\bm{r}$ and $-\frac{1}{2}\bm{r}$. The light-QCD state can be chosen to have B\nobreakdash-O quantum numbers $\lambda$ and $\eta$. In the $\bm{r}\to0$ limit, the light-QCD state is dominated by light-QCD angular-momentum/parity $j^\pi$.  We denote the state by a ket $\lvert (+,-)j^\pi, \lambda,\eta \rangle$ in the text and by a ket $\smallvket{(+,-)}{j^\pi}{\lambda}{,\eta}$ in equations. The multiplet corresponding to the $2 j + 1$ values of $\lambda$ will be labeled more concisely as $(+,-)j^\pi$. The hidden-heavy hadron state can be expressed in terms of integrals over $\bm{r}$ of these $\bm{r}$-dependent kets multiplied by wavefunctions.

The decay of a hidden-heavy hadron into a pair of heavy hadrons can proceed through the transition of a light-QCD state of the hidden-heavy hadron into a light-QCD state of a static-hadron pair with the same B\nobreakdash-O quantum numbers. Up to an appropriate normalization coefficient, the B\nobreakdash-O transition amplitude as a function of $\bm{r}$ can be defined as the time derivative of the transition amplitude between the light-QCD state $\lvert(+,-)j^\pi,\lambda,\eta\rangle$ of the hidden-heavy hadron and the light-QCD state $\bigl\lvert\bigl(j_1^{\pi_1} (+), j_2^{\pi_2} (-)\bigr)j^\prime,\lambda,\eta\bigr\rangle$ of a pair of static hadrons,
\begin{equation}
g_{\lambda,\eta}\bigl(j^\pi \to (j_1^{\pi_1}, j_2^{\pi_2}) j^\prime\bigr) \propto \frac{\mathrm{d} \hphantom{t}}{\mathrm{d} t} \vbra{\Bigl(j_1^{\pi_1} (+), j_2^{\pi_2} (-) \Bigr)}{j^\prime}{\lambda}{,\eta} \mathcal{T}_{t,t_0}  \vket{(+,-)}{j^\pi}{\lambda}{,\eta}
\label{tranrateq},
\end{equation}
with $\mathcal{T}_{t,t_0}$  the evolution operator from time $t_0$ to time $t$. The bra and the ket on the right side are functions of $\bm{r}$. The rotational symmetry of QCD implies that the transition amplitude in Eq.~\eqref{tranrateq} is a function of the distance $r=\lvert \bm{r}\rvert$ only. The subscripts on $g$ are quantum numbers $\lambda$ and $\eta$ that are conserved in light QCD with static $\bm{3}$ and $\bm{3}^\ast$ sources.

The B\nobreakdash-O transition amplitude in Eq.~\eqref{tranrateq} induces decays of a hidden-heavy hadron into pairs of heavy hadrons. The decay widths can be calculated by solving a Schr\"odinger equation with coupled hidden-heavy-hadron and heavy-hadron-pair channels. The total angular momentum of the hidden-heavy system is
\begin{equation}
\bm{J}=\bm{S}_Q + \bm{J}_\textup{light} + \bm{L}_Q,
\end{equation}
with $\bm{L}_Q$ the orbital-angular-momentum vector of $Q$ and $\bar{Q}$. As shown in Ref.~\cite{Bru23a} using the diabatic B\nobreakdash-O framework, including all the appropriate coupled channels automatically ensures conservation of total angular momentum and parity. The heavy (anti)quark spins are conserved up to corrections suppressed by $1/m_Q$. Since the total-heavy-spin vector $\bm{S}_Q$ is conserved in the heavy-quark limit, the \emph{Born\nobreakdash-Oppenheimer angular-momentum} vector
\begin{equation}
\bm{L} \equiv  \bm{J}_\textup{light} + \bm{L}_Q
\label{leq}
\end{equation}
is also conserved in the heavy-quark limit. We denote the quantum numbers for the B\nobreakdash-O angular-momentum vector $\bm{L}$ by $(L, m_L)$.  Hidden-heavy hadrons form approximately degenerate B\nobreakdash-O multiplets labeled by the angular-momentum/parity $L^P$ and other quantum numbers. We denote the quantum number for the orbital-angular-momentum vector $\bm{L}_Q$ by $L_Q$. In general, the orbital angular momentum is not conserved in the heavy-quark limit.

The hadron-pair states on both sides of Eq.~\eqref{wignereq} depend on the vector $\bm{r}$. They can be expanded in terms of orbital-angular-momentum states $\lvert L_Q^\prime, m^\prime \rangle$ and then projected onto total-angular-momentum states $\lvert J, m_J\rangle$.  Furthermore, the states on the right side of the equation can be decomposed into states with B\nobreakdash-O angular momentum $L$ using the angular-momentum recoupling identity
\begin{equation}
\vket{\Bigl[\Bigl(S_Q, j^\prime\Bigr)S,L_Q^\prime\Bigr]}{J}{m_J}{} = (-1)^{S_Q + j^\prime + L_Q^\prime + J} \sqrt{\tilde{S}} \sum_{L}  \sqrt{\tilde{L}}
\begin{Bmatrix}
S_Q	& j^\prime 	& S \\
L_Q^\prime & J 	& L \\
\end{Bmatrix} \vket{\Bigl[S_Q,\Bigl( j^\prime,L_Q^\prime\Bigr)L\Bigr]}{J}{m_J}{},
\end{equation}
where $\left\{\begin{smallmatrix} j_1 & j_2 & j_3 \\ j_4 & j_5 & j_6\end{smallmatrix}\right\}$ is a Wigner 6-$j$ symbol. Finally, the states can be projected onto states with definite $\mathcal{C}\mathcal{P}$-parity $\eta^\prime$:
\begin{align}
\vket{\Bigl\{\Bigl[\Bigl(\tfrac{1}{2}^+, j_1^{\pi_1}\Bigr) J_1, \Bigl(\tfrac{1}{2}^-, j_2^{\pi_2}\Bigr) J_2 \Bigr]S,L_Q^\prime\Bigr\}&}{J}{m_J}{,\eta^\prime}
= N (-1)^{2 j_1 + L_Q^\prime + J} \sqrt{\tilde{J}_1\tilde{J}_2  \tilde{S}} \notag \\
&\times \sum_{j^\prime, S_Q, L} 
(-1)^{j^\prime + S_Q}
\sqrt{\tilde{\jmath}^\prime \tilde{S}_Q \tilde{L}}
\begin{Bmatrix}
S_Q  & j^\prime & S \\
L_Q^\prime & J & L \\
\end{Bmatrix}
\begin{Bmatrix}
\frac{1}{2} & \frac{1}{2} & S_Q \\
j_1 & j_2 & j^\prime \\
J_1 & J_2 & S \\
\end{Bmatrix} \notag \\
&\times \vket{\Bigl\{\Bigl(\tfrac{1}{2}^+, \tfrac{1}{2}^-\Bigr) S_Q,\Bigl[\Bigl(j_1^{\pi_1}, j_2^{\pi_2} \Bigr) j^\prime,L_Q^\prime\Bigr]L\Bigr\}}{J}{m_J}{,\eta^\prime},
\label{intermediateq}
\end{align}
where the normalization coefficient $N$ is 1 unless the two static hadrons $j_1^{\pi_1}$ and $j_2^{\pi_2}$ are charge conjugates.  Even in this case, $N$ differs from 1 only if $J_1\neq J_2$, in which case $N=\sqrt{2}$. This dependence of $N$ on the quantum numbers has been suppressed in Eq.~\eqref{intermediateq}.  Note that we have dropped the arguments $(+)$ and $(-)$ since we have expanded the state of two heavy hadrons at the positions $+\frac{1}{2}\bm{r}$ and $-\frac{1}{2}\bm{r}$ into states with orbital angular momentum $L_Q^\prime$ and distance $r$.

In Ref.~\cite{Bru23a}, the diabatic B\nobreakdash-O approximation was used to derive a simple expression for the coupling potentials with total angular momentum $J$ in terms of the light-QCD transition amplitudes. Using the techniques of Ref.~\cite{Bru23a}, it is possible to derive a similar expression for the transition amplitudes with B\nobreakdash-O angular momentum $L$ in terms of the light-QCD transition amplitudes:
\begin{multline}
G_{L,\eta}^P\bigl(j^\pi, L_Q \to (j_1^{\pi_1}, j_2^{\pi_2}) j^\prime, L_Q^\prime\bigr) =
(-1)^{L_Q + L_Q^\prime} \\
\times\sum_{\lambda}
\cg{j}{\lambda}{L}{-\lambda}{L_Q}{0}
\cg{j^\prime}{\lambda}{L}{-\lambda}{L_Q^\prime}{0}
g_{\lambda,\eta}\bigl(j^\pi \to (j_1^{\pi_1}, j_2^{\pi_2}) j^\prime\bigr),
\label{mixpot1}
\end{multline}
with $L_Q$ and $L_Q^\prime$ the orbital angular momenta for the hidden-heavy hadron and the heavy-hadron pair, respectively. The superscript on $G$ is the quantum number $P$ that is exactly conserved. The subscripts on $G$ are quantum numbers $L$ and $\eta$ that are conserved in the heavy-quark limit. Note that the quantum numbers $P$, $\pi$, and $L_Q$ are constrained according to
\begin{equation}
P = \pi (-1)^{L_Q+1}.
\end{equation}
The right side is the product of the parity quantum numbers for the light-QCD state and the $Q\bar{Q}$ total-angular-momentum state.

In contrast to the coupling potentials in Ref.~\cite{Bru23a}, the transition amplitudes in Eq.~\eqref{mixpot1} do not take into account the heavy (anti)quark spins but they have the advantage that they allow the conservation of $L$ to be exploited. The corresponding transition operator can be obtained by multiplying Eq.~\eqref{mixpot1} on the left by the ket $\bigl\lvert \bigl[\bigl(j_1^{\pi_1}, j_2^{\pi_2} \bigr)j^\prime, L_Q^\prime\bigr]L, m_L, \eta \bigr\rangle$ and on the right by the bra $\bigl\langle \bigl(j^\pi, L_Q\bigr)L , m_L,\eta\bigr\rvert$ and then summing over $j^\prime$, $L_Q^\prime$, $L_Q$, $L$, $m_L$, and $\eta$. Note that we have dropped the argument $(+,-)$ of $j^\pi$ since we have expanded the state of $Q$ and $\bar{Q}$ at the positions $+\frac{1}{2}\bm{r}$ and $-\frac{1}{2}\bm{r}$ into states with orbital angular momentum $L_Q$ and distance $r$.

The transition operator with the heavy (anti)quark spins included can be obtained by multiplying the transition operator corresponding to the transition amplitudes in Eq.~\eqref{mixpot1} by the identity operator for the heavy (anti)quark spins, which can be expressed as a sum of projectors $\lvert S_Q, m\rangle\langle S_Q, m \rvert$ over the quantum numbers $S_Q$ and $m$. The coupling potentials are obtained by taking the matrix element between the bra corresponding to the heavy-hadron-pair state in Eq.~\eqref{intermediateq} and the hidden-heavy-hadron state $\Bigl\lvert\Bigl[\Bigl(\tfrac{1}{2}^+, \tfrac{1}{2}^-\Bigr) S_Q, \Bigl(j^\pi, L_Q\Bigr) L \Bigr] J, m_J, \eta^\prime \Bigr\rangle$.  Note that the state in Eq.~\eqref{intermediateq} contains a sum over the quantum numbers $j^\prime$, $S_Q$, and $L$. The matrix element simplifies in the heavy-quark limit, since $S_Q$ and $L$ are conserved in this limit. Thus only the sum over $j^\prime$ survives in the heavy-quark limit of the matrix element. The resulting coupling potentials reduce to
\begin{multline}
V^{J,P,\eta^\prime}_{S_Q,L,\eta}\Bigl(j^\pi, L_Q \to \bigl[\bigl(\tfrac{1}{2}^+, j_1^{\pi_1}\bigr) J_1, \bigl(\tfrac{1}{2}^-, j_2^{\pi_2}\bigr) J_2\bigr] S, L_Q^\prime\Bigr) = N (-1)^{2 j_1 + S_Q + L_Q^\prime + J} \sqrt{\tilde{J}_1\tilde{J}_2 \tilde{S} \tilde{S}_Q \tilde{L}} \\
\times \sum_{j^\prime} 
(-1)^{j^\prime}
\sqrt{\tilde{\jmath}^\prime}
\begin{Bmatrix}
S_Q  & j^\prime & S \\
L_Q^\prime & J & L \\
\end{Bmatrix}
\begin{Bmatrix}
\frac{1}{2} & \frac{1}{2} & S_Q \\
j_1 & j_2 & j^\prime \\
J_1 & J_2 & S \\
\end{Bmatrix}
G_{L,\eta}^P\bigl(j^\pi, L_Q \to (j_1^{\pi_1}, j_2^{\pi_2}) j^\prime, L_Q^\prime\bigr).
\label{mixpot2}
\end{multline}
The superscripts on $V$ are quantum numbers $J$, $P$, and $\eta^\prime$ that are exactly conserved. The subscripts on $V$ are quantum numbers $S_Q$, $L$, and $\eta$ that are conserved in the heavy-quark limit. Note that the quantum numbers $\eta^\prime$, $\eta$, and $S_Q$ are constrained according to
\begin{equation}
\eta^\prime=\eta(-1)^{S_Q + 1}.
\label{cpeq}
\end{equation}
The right side is the product of the $\mathcal{C}\mathcal{P}$ quantum numbers for the light-QCD state and the $Q\bar{Q}$ total-heavy-spin state.

\subsection{Selection Rules}
\label{selrulesec}

The transition of the light-QCD state $(+,-)j^\pi$ of a hidden-heavy hadron into the state of a pair of static hadrons $j_1^{\pi_1}(+)$ and $j_2^{\pi_2}(-)$ is forbidden if the B\nobreakdash-O quantum numbers of the former, $\lambda$ and $\eta$ (and $\epsilon$), cannot be obtained from any of the static-hadron-pair states $(j_1^{\pi_1}, j_2^{\pi_2}) j^\prime$ with $j^\prime$ ranging from $\lvert j_1 - j_2 \rvert$ to $j_1 + j_2$. This implies model-independent \emph{Born-Oppenheimer selection rules}.

\subsubsection{Selection Rules for \texorpdfstring{$\lambda$}{lambda}}

There is a B\nobreakdash-O selection rule that follows from conservation of $\lambda$. It states that the value of $\lvert\lambda\rvert$ cannot exceed the maximum value of $j^\prime$ for a static-hadron-pair with quantum numbers $j_1^{\pi_1}$ and $j_2^{\pi_2}$,
\begin{equation}
\lvert\lambda\rvert\leq j_1 + j_2.
\label{boselrule}
\end{equation}

If the two static hadrons are charge conjugates, this B\nobreakdash-O selection rule becomes more restrictive upon taking into account conservation of $\eta$ in accordance with Eq.~\eqref{etalighteq}. In this case, which requires $j_2 = j_1$, the range of $j^\prime$ is from 0 to $2j_1$. The largest value $j^\prime=2 j_1$ is possible only if $\eta=+1$ since $4 j_1$ is always even. If $\eta=-1$, the largest value of $j^\prime$ is $2 j_1 - 1$ and the selection rule for $\lambda$ in Eq.~\eqref{boselrule} is replaced by%
\footnote{Note that if $j_1=j_2=0$, Eq.~\eqref{boselrule} becomes $\lambda=0$ and Eq.~\eqref{specialboselrule} does not apply.}
\begin{equation}
\lvert\lambda\rvert \leq 2 j_1 - 1.
\label{specialboselrule}
\end{equation}

\subsubsection{Selection Rules for \texorpdfstring{$\epsilon$}{epsilon}}

In the special case $\lambda=0$, the B\nobreakdash-O quantum numbers are $\Sigma_\eta^\epsilon$ and there is a B\nobreakdash-O selection rule associated with conservation of $\epsilon$. Following Eq.~\eqref{refleq}, the static-hadron-pair states with integer $j^\prime$ and $\lambda=0$ are eigenstates of reflections $\mathcal{R}$ with eigenvalue $\pi_1 \pi_2(-1)^{j^\prime}$. Thus, if the light-QCD state of the hidden-heavy hadron is $\Sigma_\eta^\epsilon$, $\epsilon$ must satisfy
\begin{equation}
\epsilon = \pi_1 \pi_2(-1)^{j^\prime}.
\label{epsilonlighteq}
\end{equation}

If the two static hadrons are charge conjugates, the selection rule for $\epsilon$ in Eq.~\eqref{epsilonlighteq} implies a selection rule for $\eta$. Conservation of $\eta$, Eq.~\eqref{etalighteq}, and conservation of $\epsilon$, Eq.~\eqref{epsilonlighteq} with $\pi_2=\pi_1(-1)^{2 j_1}$, together imply
\begin{equation}
\eta = \epsilon.
\label{specialselqrule}
\end{equation}
Concretely, Eq.~\eqref{specialselqrule} states that a hidden-heavy hadron with B\nobreakdash-O quantum numbers $\Sigma_g^-$ or $\Sigma_u^+$ cannot decay into a pair of charge-conjugate static hadrons.

\subsubsection{Selection Rules for Orbital Angular Momentum and Spin}

There are selection rules associated with conservation of parity and angular momentum. For a hidden-heavy hadron with parity $P$, conservation of parity implies
\begin{equation}
P = P_1 P_2 (-1)^{L_Q^\prime},
\label{pselrule}
\end{equation}
with $P_1$ and $P_2$ the parities of the two heavy hadrons. Conservation of angular momentum implies triangle conditions that can be read from the Wigner 6-$j$ and 9-$j$ symbols in Eq.~\eqref{mixpot2}. The most interesting selection rules of this kind are triangle conditions for the triads of quantum numbers $(S_Q, j^\prime, S)$ and $(L_Q^\prime,j^\prime,L)$. They can be expressed as a selection rule on $S$:
\begin{equation}
\lvert S_Q - j^\prime\rvert \leq S \leq S_Q + j^\prime,
\label{srule}
\end{equation}
and a selection rule on $L_Q^\prime$:
\begin{equation}
\lvert L - j^\prime \rvert \leq L_Q^\prime \leq L + j^\prime.
\label{lqrule}
\end{equation}

\subsection{Relative Partial Decay Rates}
\label{branratiosec}

The coupling potentials defined by inserting Eq.~\eqref{mixpot1} into Eq.~\eqref{mixpot2} depend on the $Q\bar{Q}$ distance $r$ only through the B\nobreakdash-O transition amplitudes $g_{\lambda,\eta}$ introduced in Eq.~\eqref{tranrateq}. In general, calculating the decay width of a hidden-heavy hadron into a pair of heavy hadrons requires the solution of a coupled-channel Schr\"odinger equation. However, in some simple cases one can derive analytic expressions for the relative partial decay rates without solving any Schr\"odinger equation.

Let us consider the decays of two hidden-heavy hadrons $\Psi_{(L^P,S_{Q\alpha})J_\alpha}$ and $\Psi_{(L^P,S_{Q\beta})J_\beta}$ into heavy-hadron pairs $\Phi_{j_1^{\pi_1},J_{1\alpha}},\Phi_{j_2^{\pi_2},J_{2\alpha}}$ and $\Phi_{j_1^{\pi_1},J_{1\beta}},\Phi_{j_2^{\pi_2},J_{2\beta}}$, respectively. The hidden-heavy hadrons $\Psi_{(L^P,S_{Q\alpha})J_\alpha}$ and $\Psi_{(L^P,S_{Q\beta})J_\beta}$ belong to the same B\nobreakdash-O multiplet with angular-momentum/parity $L^P$ and they have total heavy spins $S_{Q\alpha}$ and $S_{Q\beta}$ and total hadron spins $J_\alpha$ and $J_\beta$. The heavy hadrons $\Phi_{j_1^{\pi_1},J_{1\alpha}}$ and $\Phi_{j_1^{\pi_1},J_{1\beta}}$ belong to the same HQSS doublet with light-QCD quantum numbers $j_1^{\pi_1}$ and they have spins $J_{1\alpha}$ and $J_{1\beta}$. The heavy hadrons $\Phi_{j_2^{\pi_2},J_{2\alpha}}$ and $\Phi_{j_2^{\pi_2},J_{2\beta}}$ belong to the same HQSS doublet with light-QCD quantum numbers $j_2^{\pi_2}$ and they have spins $J_{2\alpha}$ and $J_{2\beta}$.

We consider decays into heavy-hadron pairs $\Phi_{j_1^{\pi_1},J_{1\alpha}}, \Phi_{j_2^{\pi_2},J_{2\alpha}}$ and $\Phi_{j_1^{\pi_1},J_{1\beta}}, \Phi_{j_2^{\pi_2},J_{2\beta}}$ with the same orbital angular momentum $L_Q^\prime$. Their coupling potentials with the hidden-heavy hadrons $\Psi_{(L^P,S_{Q\alpha})J_\alpha}$ and $\Psi_{(L^P,S_{Q\beta})J_\beta}$ are obtained by inserting Eq.~\eqref{mixpot1} into Eq.~\eqref{mixpot2}. Note that, in addition to the orbital angular momentum $L_Q^\prime$, the coupling potentials also depend on the static angular momentum $S$, which for the heavy-hadron pairs $\Phi_{j_1^{\pi_1},J_{1\alpha}}, \Phi_{j_2^{\pi_2},J_{2\alpha}}$ or $\Phi_{j_1^{\pi_1},J_{1\beta}}, \Phi_{j_2^{\pi_2},J_{2\beta}}$ is just the total heavy-hadron spin $S_\alpha$ or $S_\beta$. We will assume here that the angular distributions of the heavy-hadron pairs $\Phi_{j_1^{\pi_1},J_{1\alpha}},\Phi_{j_2^{\pi_2},J_{2\alpha}}$ and $\Phi_{j_1^{\pi_1},J_{1\beta}},\Phi_{j_2^{\pi_2},J_{2\beta}}$ are not measured. We therefore consider decay rates summed over all possible values of $S_\alpha$ and $S_\beta$.

The phase space for a pair of heavy hadrons $\Phi_{j_1^{\pi_1},J_{1\alpha}}, \Phi_{j_2^{\pi_2},J_{2\alpha}}$ with orbital angular momentum $L_Q^\prime$ is proportional to $v_\alpha^{2 L_Q^\prime + 1}$ where $v_\alpha$ is the velocity of either heavy hadron in the center-of-momentum frame. The velocity $v_\alpha$ is determined by the mass of $\Psi_{(L^P,S_{Q\alpha})J_\alpha}$ and the masses of $\Phi_{j_1^{\pi_1},J_{1\alpha}}$ and $\Phi_{j_2^{\pi_2},J_{2\alpha}}$.

If a single B\nobreakdash-O transition amplitude $g_{\lambda,\eta}$ dominates the coupling potentials, then the sum over $\lambda$ in Eq.~\eqref{mixpot1} and the sum over $j^\prime$ in Eq.~\eqref{mixpot2} both reduce to one single term. In this case, the radial dependence factors out of the coupling potentials between $\Psi_{(L^P,S_{Q\alpha})J_\alpha}$ and $\Phi_{j_1^{\pi_1},J_{1\alpha}},\Phi_{j_2^{\pi_2},J_{2\alpha}}$ and between $\Psi_{(L^P,S_{Q\beta})J_\beta}$ and $\Phi_{j_1^{\pi_1},J_{1\beta}}, \Phi_{j_2^{\pi_2},J_{2\beta}}$. The ratio of the coupling potentials can then be expressed in terms of Wigner 6-$j$ and 9-$j$ symbols.

The simple ratio of coupling potentials implies a simple ratio of partial decay rates if in addition the kinetic energies of the heavy hadrons $\Phi_{j_1^{\pi_1}, J_{1\alpha}}$, $\Phi_{j_2^{\pi_2}, J_{2\alpha}}$ and $\Phi_{j_1^{\pi_1},J_{1\beta}}$, $\Phi_{j_2^{\pi_2},J_{2\beta}}$ are much larger than their spin splittings, that is, if the masses of the hidden-heavy hadrons $\Psi_{(L^P,S_{Q\alpha})J_\alpha}$ and $\Psi_{(L^P,S_{Q\beta})J_\beta}$ are both well above the corresponding hadron-pair thresholds. In this case, the ratio of the partial decay rates normalized by the phase-space factors $v_\alpha^{2 L_Q^\prime + 1}$ and $v_\beta^{2 L_Q^\prime + 1}$ can be approximated by the square of the ratio of the coupling potentials, which is a rational number:
\begin{multline}
\frac{\mathrm{d}\Gamma\bigl(\Psi_{(L^P,S_{Q\alpha})J_\alpha}\to (\Phi_{j_1^{\pi_1},J_{1\alpha}}, \Phi_{j_2^{\pi_2},J_{2\alpha}})_{L_Q^\prime}\bigr) / v_\alpha^{2 L_Q^\prime + 1}}{\mathrm{d}\Gamma\bigl(\Psi_{(L^P,S_{Q\beta})J_\beta}\to (\Phi_{j_1^{\pi_1},J_{1\beta}}, \Phi_{j_2^{\pi_2},J_{2\beta}})_{L_Q^\prime}\bigr) / v_\beta^{2 L_Q^\prime + 1}} \\
\approx
\frac{N_\alpha^2 \tilde{S}_{Q\alpha} \tilde{J}_{1\alpha}\tilde{J}_{2\alpha}}{N_\beta^2 \tilde{S}_{Q\beta} \tilde{J}_{1\beta}\tilde{J}_{2\beta}}
\frac{
	\displaystyle\sum_{S_\alpha} \tilde{S}_\alpha
	\begin{Bmatrix}
	S_{Q\alpha}  & j^\prime & S_\alpha \\
	L_Q^\prime & J_\alpha & L \\
	\end{Bmatrix}^2
	\begin{Bmatrix}
	\frac{1}{2} & \frac{1}{2} & S_Q \\
	j_1 & j_2 & j^\prime \\
	J_{1\alpha} & J_{2\alpha} & S_\alpha \\
	\end{Bmatrix}^2
}
{
	\displaystyle\sum_{S_\beta}
	\tilde{S}_\beta 
	\begin{Bmatrix}
	S_{Q\beta}  & j^\prime & S_\beta \\
	L_Q^\prime & J_\beta & L \\
	\end{Bmatrix}^2
	\begin{Bmatrix}
	\frac{1}{2} & \frac{1}{2} & S_Q \\
	j_1 & j_2 & j^\prime \\
	J_{1\beta} & J_{2\beta} & S_\beta \\
	\end{Bmatrix}^2
},
\label{partdecays}
\end{multline}
where $N_\alpha$ and $N_\beta$ are normalization coefficients defined after Eq.~\eqref{mixpot2}. The factor $N_\alpha^2$ is 1 unless the two static hadrons $j_1^{\pi_1}$ and $j_2^{\pi_2}$ are charge conjugates.  Even in this case, $N_\alpha^2$ differs from 1 only if $J_{1\alpha}\neq J_{2\alpha}$, in which case $N_\alpha^2=2$. If the two hidden-heavy hadrons are both $\Psi_{(L^P,S_{Q\alpha})J_\alpha}$, which requires $S_{Q\beta}=S_{Q\alpha}$ and $J_\beta=J_\alpha$,  the ratio in Eq.~\eqref{partdecays} gives a branching ratio for the hidden-heavy hadron.

Note that the relative partial decay rates in Eq.~\eqref{partdecays} are independent of the orbital angular momentum $L_Q$ of the $Q\bar{Q}$ pair inside the hidden-heavy hadrons $\Psi_{(L^P,S_{Q\alpha})J_\alpha}$ and $\Psi_{(L^P,S_{Q\beta})J_\beta}$. The dependence of the coupling potential $V^{J,P,\eta^\prime}_{S_Q,L,\eta}$ in Eq.~\eqref{mixpot2} on $L_Q$ enters only through the $G_{L,\eta}^P$ factor defined in Eq.~\eqref{mixpot1}. If only a single $g_{\lambda,\eta}$ factor contributes to the decay, then only a single $G_{L,\eta}^P$ contributes and it cancels in the relative partial decay rates. Not depending on $L_Q$ is an essential condition for the relative partial decay rates to be predictable without knowing the radial dependence of the B\nobreakdash-O potentials. In fact, a hidden-heavy-hadron state may contain several components with different values of $L_Q$ since it is in general not conserved. The relative partial decay rates in Eq.~\eqref{partdecays} do depend on the B\nobreakdash-O angular-momentum $L$, which is conserved in the heavy-quark limit.

Heavy-quark spin symmetry can be applied to heavy-hadron pairs with equal velocities, $v_\alpha=v_\beta$. Note that, because of spin splittings, the prediction of heavy-quark spin symmetry are good approximations only when the difference between the velocities $v_\alpha$ and $v_\beta$ is small.

The relative partial decay rates in Eq.~\eqref{partdecays} apply to decays in which there is one single B\nobreakdash-O transition amplitude $g_{\lambda,\eta}$. In instances where there is a dominant B\nobreakdash-O transition amplitude $g_{\lambda,\eta}$ but there are also subdominant B\nobreakdash-O transition amplitudes, the relative partial decay rates in Eq.~\eqref{partdecays} correspond to the leading-order approximation in which the subdominant B\nobreakdash-O transition amplitudes are taken to zero. Ideally, small corrections to this approximation could be implemented using perturbation theory. But these corrections would spoil the factorization of the radial dependence that is necessary to predict the relative partial decay rates without solving a Schr\"odinger equation, making such a perturbative expansion of little practical utility.

\section{Quarkonia and Quarkonium Hybrids}
\label{applicationsec}

In this section, we explicitly point out selection rules and relative partial decay rates for decays of quarkonia and quarkonium hybrids into pairs of heavy mesons. We focus on the three lightest HQSS doublets of heavy mesons, that is, $S$-wave, $P^-$-wave, and $P^+$-wave mesons. The light-QCD quantum numbers of an $S$-wave meson are $\frac{1}{2}^{-}$ if it has the flavor of a light antiquark and $\frac{1}{2}^{+}$ if it has the flavor of a light quark. The corresponding light-QCD quantum numbers are $\frac{1}{2}^{+}$ and $\frac{1}{2}^{-}$ for a $P^-$-wave meson and $\frac{3}{2}^{+}$ and $\frac{3}{2}^{-}$ for a $P^+$-wave meson. The $J^P$ quantum numbers of an $S$-wave meson are $0^-$ or $1^-$. They are $0^+$ or $1^+$ for a $P^-$-wave meson and $1^+$ or $2^+$ for a $P^+$-wave meson.

We will only consider decays into the three lightest pairs of heavy mesons: $S+S$, $S+P^-$, and $S+P^+$. For the sake of simplicity, we choose the first heavy meson ($S$-wave) to have the flavor of a light antiquark and the second heavy meson ($S$-, $P^-$-, or $P^+$-wave) to have the flavor of a light quark. The $\mathcal{C}\mathcal{P}$ eigenvalue $\eta^\prime$ of the meson pair is determined by Eq.~\eqref{cpeq} in terms of the B\nobreakdash-O quantum number $\eta$ of the hidden-heavy hadron and the total heavy spin $S_Q$.

\subsection{Quarkonium Decays}
\label{quarkoniumsec}

Quarkonia are associated with bound states in the lowest confining $\Sigma_g^+$ potential, with B\nobreakdash-O quantum numbers $\lambda=0$, $\eta=+1$, and $\epsilon=+1$. Their light-QCD state at short distances is dominated by angular-momentum/parity $j^\pi=0^+$. The B\nobreakdash-O angular momentum is therefore equal to the orbital angular momentum, $L=L_Q$. A quarkonium multiplet consists of several quarkonia with same orbital angular momentum $L_Q$ and parity $P=(-1)^{L_Q+1}$ but different quantum numbers $S_Q$ and $J$. The $\mathcal{C}$-parity is determined by the orbital angular momentum $L_Q$ and the total heavy spin $S_Q$ as $C=(-1)^{L_Q + S_Q}$.

\subsubsection{Selection Rules}

An $S+S$ meson pair corresponds to light-QCD quantum numbers $\bigl(\frac{1}{2}^-,\frac{1}{2}^+\bigr) j^\prime$ with $j^\prime=0$ or $1$. The selection rule for $\epsilon$ in Eq.~\eqref{epsilonlighteq} requires $j^\prime=1$. The parities of the mesons are $P_1=P_2=-1$. Conservation of parity in Eq.~\eqref{pselrule} requires $(-1)^{L_Q+1}=(-1)^{L_Q^\prime}$. The selection rule for $L_Q^\prime$ in Eq.~\eqref{lqrule} then requires
\begin{equation}
L_Q^\prime =
\begin{cases}
L_Q - 1, L_Q+1 & \text{if $L_Q \geq 1$,} \\
1 & \text{if $L_Q =0$.}
\end{cases}
\label{qselruless}
\end{equation}
This selection rule was derived long ago using quark-pair-creation models; see, for instance, Refs.~\cite{LeY88,Bur14} and references therein. It implies that $S+S$ meson pairs in a relative $S$-wave ($L_Q^\prime=0$) can only be produced by the decay of quarkonia in multiplets with $L_Q=1$.

An $S+P^-$ meson pair corresponds to light-QCD quantum numbers $\bigl(\frac{1}{2}^-,\frac{1}{2}^-\bigr) j^\prime$ with $j^\prime=0$ or $1$. The selection rule for $\epsilon$ in Eq.~\eqref{epsilonlighteq} requires $j^\prime=0$. The selection rule for $S$ in Eq.~\eqref{srule} then requires
\begin{equation}
S=S_Q.
\label{qselrulespm}
\end{equation}
The selection rule for $L_Q^\prime$ in Eq.~\eqref{lqrule} requires
\begin{equation}
L_Q^\prime=L_Q.
\end{equation}
This selection rule implies that $S+P^-$ meson pairs in a relative $S$-wave can only be produced by the decay of quarkonia in multiplets with $L_Q=0$.

An $S+P^+$ meson pair corresponds to light-QCD quantum numbers $\bigl(\frac{1}{2}^-,\frac{3}{2}^-\bigr) j^\prime$ with $j^\prime=1$ or $2$. The selection rule for $\epsilon$ in Eq.~\eqref{epsilonlighteq} requires $j^\prime=2$. The parities of the mesons are $P_1=-1$, $P_2=+1$. Conservation of parity in Eq.~\eqref{pselrule} requires $(-1)^{L_Q+1}=(-1)^{L_Q^\prime+1}$. The selection rule for $L_Q^\prime$ in Eq.~\eqref{lqrule} then requires
\begin{equation}
L_Q^\prime=
\begin{cases}
L_Q - 2, L_Q, L_Q + 2 & \text{if $L_Q\geq2$,} \\
1, 3 & \text{if $L_Q=1$,} \\
2 & \text{if $L_Q=0$.}
\end{cases}
\label{qselrulespp}
\end{equation}
This selection rule implies that $S+P^+$ meson pairs in a relative $S$-wave can only be produced by the decay of quarkonia in multiplets with $L_Q=2$. Conversely, for the decays of $S$-wave quarkonia ($L_Q=0$), whose $J^{PC}$ quantum numbers are either $0^{-+}$ or $1^{--}$, this selection rule requires $L_Q^\prime=2$. This selection rule for the $1^{--}$ case was previously derived by Li and Voloshin, who pointed out that the $S$-wave production in $e^+e^-$ annihilation of an $S+P^+$ meson pair is suppressed \cite{Vol13}.

\subsubsection{Relative Partial Decay Rates}

For heavy-meson-pair decays of quarkonia, which are associated with bound states in the $\Sigma_g^+$ potential, conservation of $\lambda$ requires $\lambda=0$. Therefore, the relative partial decay rates for quarkonia in the same B\nobreakdash-O multiplet into pairs of heavy mesons in specific HQSS doublets are predictable without solving any Schr\"odinger equation if one value of $j^\prime$ dominates the transition. As shown before Eqs.~\eqref{qselruless}, \eqref{qselrulespm}, and \eqref{qselrulespp}, this condition is fulfilled by $S+S$, $S+P^-$, and $S+P^+$ meson pairs. The dominant values of $j^\prime$ are 1, 0, and 2, respectively. Here we will limit ourselves to quarkonium multiplets with $L_Q=0$ and 1, which are the quantum numbers of the lowest two quarkonium multiplets.

A quarkonium multiplet with orbital angular momentum $L_Q=0$ ($S$-wave) has angular-momentum/parity $L^P=0^-$. A multiplet with these quantum numbers consists of one $S_Q=0$ state with $J^{PC}=0^{-+}$ and one $S_Q=1$ state with $J^{PC}=1^{--}$. Our predictions for the relative partial decay rates into $S+S$, $S+P^-$, and $S+P^+$ meson pairs are reported in Table~\ref{quarkoniuml0} in the form of ratios of integers.

\begin{table}
\caption{\label{quarkoniuml0}Relative partial decay rates for $S$-wave quarkonia ($L_Q=0$) with $J^{PC}=0^{-+}$ and $1^{--}$ into $S+S$, $S+P^-$, and $S+P^+$ meson pairs.}
\begin{ruledtabular}
\begin{tabular}{lcrr}
& & $0^{-+}$	& $1^{--}$	\\
\hline
\multirow{3}{*}{$S+S $ ($P$-wave)}	& $B\bar{B}$ 				& 0	& 1	\\
							& $B\bar{B}^\ast+B^\ast\bar{B}$ 	& 6	& 4	\\
							& $B^\ast\bar{B}^\ast$			& 6	& 7	\\
\hline
\multirow{4}{*}{$S+P^-$ ($S$-wave)}	& $B^\ast_0\bar{B}+B\bar{B}^\ast_0$		& 1	& 0	\\
							& $B^\ast_0\bar{B}^\ast+B^\ast\bar{B}^\ast_0$	& 0	& 1	\\
							& $B_1\bar{B}+B\bar{B}_1$				& 0	& 1	\\
							& $B_1\bar{B}^\ast+B^\ast\bar{B}_1$		& 3	& 2	\\
\hline
\multirow{4}{*}{$S+P^+$ ($D$-wave)}	& $B_1\bar{B}+B\bar{B}_1$ 				& 0 	& 1	\\
								& $B_1\bar{B}^\ast+B^\ast\bar{B}_1$		& 3 	& 2	\\
								& $B_2^\ast\bar{B}+B\bar{B}^\ast_2$		& 2	& 1	\\
								& $B_2^\ast\bar{B}^\ast+B^\ast\bar{B}^\ast_2$	& 3	& 4	\\
\end{tabular}
\end{ruledtabular}
\end{table}

A quarkonium multiplet with orbital angular momentum $L_Q=1$ ($P$-wave) has angular-momentum/parity $L^P=1^+$. A multiplet with these quantum numbers consists of one $S_Q=0$ state with $J^{PC}=1^{+-}$ and three $S_Q=1$ states with $J^{PC}=0^{++}$, $1^{++}$, and $2^{++}$. Our predictions for the relative partial decay rates into $S+S$, $S+P^-$, and $S+P^+$ meson pairs are reported in Table~\ref{quarkoniuml1} in the form of ratios of integers.

\begin{table}
\caption{\label{quarkoniuml1}Relative partial decay rates for $P$-wave quarkonia ($L_Q=1$) with $J^{PC}=1^{+-}$, $0^{++}$, $1^{++}$, and $2^{++}$ into $S+S$, $S+P^-$, and $S+P^+$ meson pairs.}
\begin{ruledtabular}
\begin{tabular}{lcrrrr}
& & $1^{+-}$	& $0^{++}$ & $1^{++}$ & $2^{++}$	\\
\hline
\multirow{3}{*}{$S+S$ ($S$-wave)}	& $B\bar{B}$				& 0		& 3		& 0		& 0		\\
							& $B\bar{B}^\ast+B^\ast\bar{B}$ 	& 2		& 0 		& 4		& 0		\\
							& $B^\ast\bar{B}^\ast$			& 2		& 1		& 0		& 4		\\
\hline
\multirow{3}{*}{$S+S$ ($D$-wave)}	& $B\bar{B}$ 				& 0 	& 0 	& 0		& 3		\\
							& $B\bar{B}^\ast+B^\ast\bar{B}$ 	& 10	& 0	& 5		& 9		\\
							& $B^\ast\bar{B}^\ast$			& 10	& 20	& 15		& 8		\\
\hline
\multirow{4}{*}{$S+P^-$ ($P$-wave)}	& $B^\ast_0\bar{B}+B\bar{B}^\ast_0$		& 1		& 0		& 0		& 0		\\
							& $B^\ast_0\bar{B}^\ast+B^\ast\bar{B}^\ast_0$	& 0		& 1		& 1		& 1		\\
							& $B_1\bar{B}+B\bar{B}_1$				& 0		& 1		& 1		& 1		\\
							& $B_1\bar{B}^\ast+B^\ast\bar{B}_1$		& 3		& 2		& 2		& 2		\\
\hline
\multirow{4}{*}{$S+P^+$ ($P$-wave)}	& $B_1\bar{B}+B\bar{B}_1$				& 0	& 100	& 25	& 1	\\
								& $B_1\bar{B}^\ast+B^\ast\bar{B}_1$		& 60	& 50	& 80	& 14	\\
								& $B_2^\ast\bar{B}+B\bar{B}^\ast_2$		& 40	& 0	& 45	& 9	\\
								& $B_2^\ast\bar{B}^\ast+B^\ast\bar{B}^\ast_2$	& 60	& 10	& 10	& 136	\\
\hline
\multirow{4}{*}{$S+P^+$ ($F$-wave)}	& $B_1\bar{B}+B\bar{B}_1$				& 0	& 0	& 0	& 18	\\
								& $B_1\bar{B}^\ast+B^\ast\bar{B}_1$		& 30	& 0	& 15	& 27	\\
								& $B_2^\ast\bar{B}+B\bar{B}^\ast_2$		& 20	& 0	& 10	& 12	\\
								& $B_2^\ast\bar{B}^\ast+B^\ast\bar{B}^\ast_2$	& 30	& 80	& 55	& 23	\\
\end{tabular}
\end{ruledtabular}
\end{table}

Under the narrow-resonance approximation, the ratios of the cross-sections for $e^+e^-$ annihilation into pairs of heavy mesons with the same orbital angular momentum $L_Q^\prime$ at the center-of-mass energy of a $J^{PC}=1^{--}$ $Q\bar{Q}$ resonance are given by Eq.~\eqref{partdecays} by replacing $\Psi_{(L^P,S_{Q\alpha})J_\alpha}$ and $\Psi_{(L^P,S_{Q\beta})J_\beta}$ with the $Q\bar{Q}$ state with $S_Q=1$, $L=L_Q=0$, and $J=1$. The $e^+e^-$ production cross-sections for $B\bar{B}$, $B\bar{B}^\ast+B^\ast\bar{B}$, and $B^\ast\bar{B}^\ast$ are in the proportion $1:4:7$, as can be seen from the entries for $J^{PC}=1^{--}$ in the first rows of Table~\ref{quarkoniuml0}. This result was derived long ago using heavy-quark spin symmetry \cite{DeR77}. In the $B^\ast\bar{B}^\ast$ cross section, the factor of $7$ can be decomposed into the contributions $1/3$ and $20/3$ from the sum over total-heavy-meson spins $S=0$ and $S=2$ \cite{Kai03}.

\subsection{Quarkonium Hybrid Decays}
\label{hybridsec}

The lowest-energy quarkonium hybrids are associated with bound states in confining $\Pi_u$ and $\Sigma_u^-$ potentials. Their light-QCD state at short distances is dominated by angular-momentum/parity $j^\pi=1^+$. A hybrid multiplet consists of several quarkonium hybrids with the same B\nobreakdash-O angular momentum $L$ and parity $P$ but different quantum numbers $S_Q$ and $J$. The $\mathcal{C}$-parity is determined by the parity $P$ and the total heavy spin $S_Q$ as $C=P(-1)^{S_Q}$. There are three different types of hybrid multiplets:
\begin{enumerate}[i)]
\item multiplets with $L\geq1$ and $P=+(-1)^L$ consisting of bound states in coupled $\Pi_u$ and $\Sigma_u^-$ potentials;
\item multiplets with $L\geq1$ and $P=-(-1)^L$ consisting of bound states in the $\Pi_u$ potential;
\item{} multiplets with $L=0$ and $P=+1$ consisting of bound states in the $\Sigma_u^-$ potential.
\end{enumerate}

\subsubsection{Selection Rules}

An $S+S$ meson pair corresponds to light-QCD quantum numbers $\bigl(\frac{1}{2}^-,\frac{1}{2}^+\bigr) j^\prime$ with $j^\prime=0$ or $1$. Since the $\Pi_u$ and $\Sigma_u^-$ potentials have $\mathcal{C}\mathcal{P}$ quantum number $\eta=-1$, conservation of $\mathcal{C}\mathcal{P}$, Eq.~\eqref{etalighteq}, requires $j^\prime = 0$ and therefore $\lambda=0$. This implies that quarkonium hybrids associated with bound states in the $\Pi_u$ potential are forbidden to decay into $S+S$ meson pairs. On the other hand, quarkonium hybrids associated with bound states in the $\Sigma_u^-$ potential or bound states in the coupled $\Pi_u$ and $\Sigma_u^-$ potentials are allowed to decay into $S+S$ meson pairs. This model-independent result, which was first pointed out in Ref.~\cite{Bru23b}, contradicts the conventional wisdom for the last 40 years based on quark-pair-creation models that hybrid mesons are forbidden to decay into pairs of heavy-mesons with identical spatial structure \cite{Tan82,LeY84,Is85,Idd88,Ish91,Clo94,Pag97,Pag99,Kou05}. The selection rule for $S$ in Eq.~\eqref{srule} then requires
\begin{equation}
S=S_Q,
\label{hselruless}
\end{equation}
which was first derived in Ref.~\cite{Bru23b}. The spin selection rule in Eq.~\eqref{hselruless} implies that quarkonium hybrids may decay into either $B\bar{B}$, which has $S=0$, or $B\bar{B}^\ast$ and $B^\ast\bar{B}$, which have $S=1$, but never both. The selection rule for $L_Q^\prime$ in Eq.~\eqref{lqrule} requires
\begin{equation}
L_Q^\prime=L,
\end{equation}
which was first derived in Ref.~\cite{Cas24}. This selection rule implies that $S+S$ meson pairs in a relative $S$-wave ($L_Q^\prime=0$) can only be produced by the decays of quarkonium hybrids in multiplets with $L^P=0^+$.

An $S+P^-$ meson pair corresponds to light-QCD quantum numbers $\bigl(\frac{1}{2}^-,\frac{1}{2}^-\bigr) j^\prime$ with $j^\prime=0$ or $1$. By conservation of $\lambda$, the $\Pi_u$ potential couples only to $j^\prime=1$. For the $\Sigma_u^-$ potential, the selection rule for $\epsilon$ in Eq.~\eqref{epsilonlighteq} requires $j^\prime=1$ as well. Therefore, the decay of a quarkonium hybrid into an $S+P^-$ meson pair proceeds only through $j^\prime=1$. The parities of the mesons are $P_1=-1$, $P_2=+1$. Conservation of parity in Eq.~\eqref{pselrule} requires $P=(-1)^{L_Q^\prime+1}$. The selection rule for $L_Q^\prime$ in Eq.~\eqref{lqrule} then requires
\begin{equation}
L_Q^\prime=
\begin{cases}
L-1,L+1 	& \text{if $L\geq1$ and $P=+(-1)^L$,} \\
L & \text{if $L\geq1$ and $P=-(-1)^L$,} \\
1 & \text{if $L=0$ and $P=+1$.}
\end{cases}
\label{hselrulespm}
\end{equation}
This selection rule implies that $S+P^-$ meson pairs in a relative $S$-wave can only be produced by the decays of quarkonium hybrids in multiplets with $L^P=1^-$.

An $S+P^+$ meson pair corresponds to light-QCD quantum numbers $\bigl(\frac{1}{2}^-,\frac{3}{2}^-\bigr) j^\prime$ with $j^\prime=1$ or $2$. The $\Pi_u$ potential can couple to both $j^\prime=1$ and 2. For the $\Sigma_u^-$ potential, the selection rule for $\epsilon$ in Eq.~\eqref{epsilonlighteq} requires $j^\prime=1$. For decays of a quarkonium hybrid with $L=0$, which is a bound state in the $\Sigma_u^-$ potential, one has $j^\prime=1$. For decays of a quarkonium hybrid with $L\geq1$, which is a bound state either in the $\Pi_u$ potential or in coupled $\Pi_u$ and $\Sigma_u^-$ potentials, one has $j^\prime = 1$ or 2. The parities of the mesons are $P_1=-1$, $P_2=+1$. Conservation of parity in Eq.~\eqref{pselrule} requires $P=(-1)^{L_Q^\prime+1}$. The selection rule for $L_Q^\prime$ in Eq.~\eqref{lqrule} then requires
\begin{equation}
L_Q^\prime=
\begin{cases}
L-2,L,L+2 & \text{if $L\geq2$ and $P=-(-1)^L$,} \\
L-1,L+1 	& \text{if $L\geq1$ and $P=+(-1)^L$,} \\
1,3 & \text{if $L=1$ and $P=+1$,} \\
1 & \text{if $L=0$ and $P=+1$.}
\end{cases}
\label{hselrulespp}
\end{equation}
This selection rule implies that $S+P^+$ meson pairs in a relative $S$-wave can only be produced by the decays of quarkonium hybrids in multiplets with $L^P=1^-$ and $2^-$.

\subsubsection{Relative Partial Decay Rates}

Whether the relative partial decay rates for quarkonium hybrids into pairs of heavy mesons are predictable without solving any Schr\"odinger equation depends on the hybrid B\nobreakdash-O multiplet as well as the heavy-meson HQSS doublets. Here we will limit ourselves to hybrid multiplets with the same angular-momentum/parity $L^P$ as the lowest three hybrid multiplets.

The lowest hybrid multiplet has angular-momentum/parity $L^P=1^-$. A multiplet with these quantum numbers consists of one $S_Q=0$ state with $J^{PC}=1^{--}$ and three $S_Q=1$ states with $J^{PC}=0^{-+}$, $1^{-+}$, and $2^{-+}$. Since they are associated with a bound state in coupled $\Pi_u$ and $\Sigma_u^-$ potentials, the value of $\lambda$ can be either 0 or 1. Therefore, the relative partial decay rates are predictable only if $j^\prime=0$ dominates the transition. As shown before Eqs.~\eqref{hselruless}, \eqref{hselrulespm}, and \eqref{hselrulespp}, this condition is fulfilled by $S+S$ meson pairs but not by $S+P^-$ or $S+P^+$ meson pairs. Our predictions for the relative partial decay rates into $S+S$ meson pairs are reported in Table~\ref{hybridl1minus} in the form of ratios of integers. Note that these nonzero decay rates contradict the conventional wisdom on decays of hybrid mesons, as first pointed out in Ref.~\cite{Bru23b}.

\begin{table}
\caption{\label{hybridl1minus}Relative partial decay rates for $L^P=1^-$ quarkonium hybrids with $J^{PC}=1^{--}$, $0^{-+}$, $1^{-+}$, and $2^{-+}$ into $S+S$ meson pairs. Decays into $S+P^-$ and $S+P^+$ meson pairs are not listed, because the relative partial decay rates are not completely determined by B-O symmetries.}
\begin{ruledtabular}
\begin{tabular}{lcrrrr}
& & $1^{--}$	& $0^{-+}$		& $1^{-+}$	& $2^{-+}$	\\
\hline
\multirow{3}{*}{$S+S$ ($P$-wave)}	& $B\bar{B}$ 				& 1		& 0		& 0		& 0		\\
							& $B\bar{B}^\ast+B^\ast\bar{B}$	& 0		& 2		& 2		& 2		\\
							& $B^\ast\bar{B}^\ast$			& 3		& 2		& 2		& 2		\\
\end{tabular}
\end{ruledtabular}
\end{table}

The second lowest hybrid multiplet has angular-momentum/parity $L^P=1^+$. A multiplet with these quantum numbers consists of one $S_Q=0$ state with $J^{PC}=1^{++}$ and three $S_Q=1$ states with $J^{PC}=0^{+-}$, $1^{+-}$, and $2^{+-}$. Since they are associated with a bound state in the $\Pi_u$ potential, one necessarily has $\lambda=1$. Therefore, the relative partial decay rates are predictable if one value of $j^\prime$ dominates the transition. As shown before Eq.~\eqref{hselrulespm}, this condition is fulfilled by $S+P^-$ meson pairs since the dominant value of $j^\prime$ is 1. As shown before Eq.~\eqref{hselrulespp}, this condition is generally not satisfied by $S+P^+$ meson pairs since there are two dominant values of $j^\prime$, that is, 1 and 2. Incidentally, the condition is satisfied by $S+P^+$ meson pairs in a relative $F$-wave ($L_Q^\prime=3$) since a single dominant value $j^\prime=2$ is required by the selection rule in Eq.~\eqref{lqrule} with $L_Q^\prime=3$ and $L=1$, which forbids $j^\prime=1$. Decays into $S+S$ meson pairs are forbidden, since these hybrid multiplets have no component associated with the $\Sigma_u^-$ potential. Our predictions for the relative partial decay rates into $S+P^-$ and $S+P^+$ ($F$-wave) meson pairs are reported in Table~\ref{hybridl1plus} in the form of ratios of integers.

\begin{table}
\caption{\label{hybridl1plus}Relative partial decay rates for $L^P=1^+$ quarkonium hybrids with $J^{PC}=1^{++}$, $0^{+-}$, $1^{+-}$, and $2^{+-}$ into $S+P^-$ and $S+P^+$ ($F$-wave) meson pairs. Decays into $S+S$ meson pairs are not listed, because they are forbidden by B-O symmetries. Decays into $S+P^+$ ($P$-wave) meson pairs are not listed, because the relative partial decay rates are not completely determined by B-O symmetries.}
\begin{ruledtabular}
\begin{tabular}{lcrrrr}
& & $1^{++}$	& $0^{+-}$		& $1^{+-}$	& $2^{+-}$	\\
\hline
\multirow{4}{*}{$S+P^-$ ($P$-wave)}	& $B^\ast_0\bar{B}+B\bar{B}^\ast_0$		& 0		& 0		& 2		& 0		\\
							& $B^\ast_0\bar{B}^\ast+B^\ast\bar{B}^\ast_0$	& 2		& 4		& 1		& 1		\\
							& $B_1\bar{B}+B\bar{B}_1$				& 2		& 4		& 1		& 1		\\
							& $B_1\bar{B}^\ast+B^\ast\bar{B}_1$		& 4		& 0		& 4		& 6		\\
\hline
\multirow{4}{*}{$S+P^+$ ($F$-wave)}	& $B_1\bar{B}+B\bar{B}_1$ 				& 0 		& 0		& 0		& 18		\\
								& $B_1\bar{B}^\ast+B^\ast\bar{B}_1$			& 30		& 0		& 15		& 27		\\
								& $B_2^\ast\bar{B}+B\bar{B}^\ast_2$			& 20		& 0		& 10		& 12		\\
								& $B_2^\ast\bar{B}^\ast+B^\ast\bar{B}^\ast_2$	& 30		& 80		& 55		& 23		\\
\end{tabular}
\end{ruledtabular}
\end{table}

The third lowest hybrid multiplet has angular-momentum/parity $L^P=0^+$. A multiplet with these quantum numbers consists of one $S_Q=0$ state with $J^{PC}=0^{++}$ and one $S_Q=1$ state with $J^{PC}=1^{+-}$. Since they are associated with bound states in the $\Sigma_u^-$ potential, one necessarily has $\lambda=0$. Therefore, the relative partial decay rates are predictable if one $j^\prime$ dominates the transition. As shown before Eqs.~\eqref{hselruless}, \eqref{hselrulespm}, and \eqref{hselrulespp}, this condition is fulfilled by $S+S$, $S+P^-$, and $S+P^+$ meson pairs. The dominant values of $j^\prime$ are 0, 1, and 1, respectively. Our predictions for the relative partial decay rates into $S+S$, $S+P^-$, and $S+P^+$ meson pairs are reported in Table~\ref{hybridl0} in the form of ratios of integers. Note that the nonzero decay rates into $S+S$ meson pairs contradict the conventional wisdom on decays of hybrid mesons, as first pointed out in Ref.~\cite{Bru23b}.

\begin{table}
\caption{\label{hybridl0}Relative partial decay rates for $L^P=0^+$ quarkonium hybrids with $J^{PC}=0^{++}$ and $1^{+-}$ into $S+S$, $S+P^-$, and $S+P^+$ meson pairs.}
\begin{ruledtabular}
\begin{tabular}{lcrr}
& & $0^{++}$	& $1^{+-}$	\\
\hline
\multirow{3}{*}{$S+S $ ($S$-wave)}	& $B\bar{B}$ 				& 1		& 0		\\
							& $B\bar{B}^\ast+B^\ast\bar{B}$ 	& 0		& 2		\\
							& $B^\ast\bar{B}^\ast$			& 3		& 2		\\
\hline
\multirow{4}{*}{$S+P^-$ ($P$-wave)}	& $B^\ast_0\bar{B}+B\bar{B}^\ast_0$		& 0		& 1		\\
							& $B^\ast_0\bar{B}^\ast+B^\ast\bar{B}^\ast_0$	& 3		& 2		\\
							& $B_1\bar{B}+B\bar{B}_1$				& 3		& 2		\\
							& $B_1\bar{B}^\ast+B^\ast\bar{B}_1$		& 6		& 7		\\
\hline
\multirow{4}{*}{$S+P^+$ ($P$-wave)}	& $B_1\bar{B}+B\bar{B}_1$ 				& 6 		& 1		\\
								& $B_1\bar{B}^\ast+B^\ast\bar{B}_1$		& 3 		& 8		\\
								& $B_2^\ast\bar{B}+B\bar{B}^\ast_2$		& 0		& 5		\\
								& $B_2^\ast\bar{B}^\ast+B^\ast\bar{B}^\ast_2$	& 15		& 10		\\
\end{tabular}
\end{ruledtabular}
\end{table}

\section{Comparison with quark-pair-creation Models}
\label{modelsec}

In constituent-quark models, the strong decay of a quark-antiquark meson or a 3-quark baryon requires the creation of light quark-antiquark pairs. Strong decays of a hadron into pairs of hadrons have been calculated using quark-pair-creation models; see Refs.~\cite{LeY88,Bur14} and references therein. The most popular models fall under the category of \emph{nonflip triplet} decay models, which assume the transition operator has a vector structure both in coordinate space and in spin space. The dependence of the transition amplitude on angular-momentum quantum numbers is identical in all nonflip triplet decay models. The angular-momentum coefficients and selection rules for the decays of a meson into pairs of mesons have been derived in the most general form in Ref.~\cite{Bur14}. In this section, we briefly review these results and compare them with the B\nobreakdash-O predictions.

\subsection{Transition Amplitudes}

In constituent-quark models, a meson is labeled by a ``principal'' quantum number $n$ ($n=1$ for the ground state and $n=2,3,\dots$ for excited states) and by quantum numbers for the total constituent-quark spin vector $\bm{S}$, the (internal) orbital angular momentum vector $\bm{L}$, and the total angular momentum vector $\bm{J}=\bm{S}+\bm{L}$. The corresponding quantum numbers of the decaying meson are $n$, $S_Q$, $L_Q$, and $J$. The corresponding quantum numbers of each final-state meson $i=1,2$ are $n_i$, $S_i$, $L_i$, and $J_i$. We label the total spin of the two heavy mesons by $S$ and the (relative) orbital angular momentum between them by $L_Q^\prime$. In Ref.~\cite{Bur14}, the transition amplitude $M$ is generally expressed as%
\footnote{The following expressions are taken from Ref.~\cite{Bur14} with the substitutions $S\to S_Q$, $L\to L_Q$, $j\to S$, $l\to L_Q^\prime$, $l^\prime \to L^{\prime\prime}$ to facilitate comparison with the results of this paper.}
\begin{equation}
M_{S,L_Q^\prime} 
\begin{bmatrix}
n & S_Q & L_Q & J \\
n_1 & S_1 & L_1 & J_1 \\
n_2 & S_2 & L_2 & J_2
\end{bmatrix}_\pm = 
\sum_{L^\prime,L^{\prime\prime}}
\xi_{S,L_Q^\prime}^{L^\prime,L^{\prime\prime}}
\begin{bmatrix}
S_Q & L_Q & J \\
S_1 & L_1 & J_1 \\
S_2 & L_2 & J_2 \\
\end{bmatrix}_\pm
A_{L_Q^\prime}^{L^\prime,L^{\prime\prime}}
\begin{bmatrix}
n & L_Q \\
n_1 & L_1 \\
n_2 & L_ 2 \\
\end{bmatrix}_\pm,
\label{modelamp}
\end{equation}
where the $\xi$'s are angular-momentum coefficients that are the same for all nonflip triplet decay models and the $A$'s are matrix elements involving the spatial overlap of quark-model wavefunctions that depend on the specific model. Their dependence on the quantum numbers associated with the initial and final mesons is denoted by arrays of numbers in square brackets. The sums in Eq.~\eqref{modelamp} are over quantum numbers $L^\prime$ and $L^{\prime\prime}$ associated with the orbital-angular-momentum vectors $\bm{L}^\prime \equiv \bm{L}_1 + \bm{L}_2$ and $\bm{L}^{\prime\prime} \equiv \bm{L}^\prime + \bm{L}_Q^\prime$. The subscripts $\pm$ in Eq.~\eqref{modelamp} stand for two different ``topologies'' of the decay. The interference between the $(+)$ and $(-)$ topologies ensures conservation of $\mathcal{C}$-parity. The angular-momentum coefficients can be written explicitly in terms of Wigner 6-$j$ and 9-$j$ symbols as \cite{Bar08}
\begin{multline}
\xi_{S,L_Q^\prime}^{L^\prime,L^{\prime\prime}}
\begin{bmatrix}
S_Q & L_Q & J \\
S_1 & L_1 & J_1 \\
S_2 & L_2 & J_2 \\
\end{bmatrix}_\pm = (\pm1)^{S_Q + S_1 + S_2 + 1} (-1)^{L_Q^\prime+L^\prime+L^{\prime\prime} + S_Q + S_2 + 1} \sqrt{3 \tilde{S}_Q \tilde{S}_1 \tilde{S}_2 \tilde{S}\tilde{L}_Q\tilde{L}^\prime\tilde{L}^{\prime\prime}\tilde{J}_1\tilde{J}_2} \\
\times \sum_{S^\prime} (-1)^{S^\prime} S^\prime
\begin{Bmatrix}
S^\prime & L^\prime & S \\
L_Q^\prime & J & L^{\prime\prime} \\
\end{Bmatrix}
\begin{Bmatrix}
S_Q & L_Q & J \\
L^{\prime\prime} & S^\prime & 1 \\
\end{Bmatrix}
\begin{Bmatrix}
S_1 & L_1 & J_1 \\
S_2 & L_2 & J_2 \\
S^\prime & L^\prime & S \\
\end{Bmatrix}
\begin{Bmatrix}
\frac{1}{2} & \frac{1}{2} & S_1 \\
\frac{1}{2} & \frac{1}{2} & S_2 \\
S_Q & 1 & S^\prime \\
\end{Bmatrix},
\label{modelcoef}
\end{multline}
where the sum is over the quantum number $S^\prime$ for the spin vector $\bm{S}^\prime \equiv \bm{S}_1 + \bm{S}_2$.  The assumed vector structure of the transition operator in coordinate space enters through an entry 1 in one of the 6-$j$ symbols, which implies a triangle condition on the quantum numbers $L^{\prime\prime}$, $L_Q$, and 1. The assumed vector structure of the transition operator in spin space enters through an entry 1 in one of the 9-$j$ symbols, which implies a triangle condition on the quantum numbers $S^\prime$, $S_Q$, and 1.

The transition amplitude in Eq.~\eqref{modelamp} applies to ordinary quark-antiquark mesons. There is a simple generalization that applies also to hybrid mesons in some but not all nonflip triplet decay models; see Ref.~\cite{Bur14} and references therein. In flux-tube models, the same angular-momentum coefficients $\xi$ in Eq.~\eqref{modelcoef} apply to the decays of hybrid mesons into meson pairs, with the substitution of $L_Q$ by an angular-momentum $L$ associated with the sum of the quark and flux-tube angular momenta. In constituent-gluon models, the angular-momentum coefficients are generally different but they coincide with Eq.~\eqref{modelcoef} in simple cases.

\subsection{Selection Rules}

The transition amplitudes for the decays of a meson into pairs of mesons are obtained by inserting Eq.~\eqref{modelcoef} into Eq.~\eqref{modelamp}. Zeroes of the transition amplitude correspond to selection rules. Zeroes of the angular-momentum coefficients $\xi$ correspond to selection rules associated with conservation of angular momentum. Zeroes of the spatial matrix elements $A$ correspond to selection rules associated with conservation of parity. These selection rules are described in Ref.~\cite{Bur14}.

For decays into meson pairs with identical spatial wavefunctions, there is an additional selection rule that corresponds to a cancellation in the sum of the two topologies $(+)$ and $(-)$ in Eq.~\eqref{modelamp}. It was derived in its most general form in Ref.~\cite{Pag97} using the symmetries of the matrix elements $A$ for nonrelativistic quark models. The rule states that a quark-antiquark meson is allowed to decay into pairs of mesons with identical spatial wavefunctions only if the $\mathcal{C}\mathcal{P}$-parity of the decaying meson and its total constituent-quark spin $S_Q$ stand in the relation $CP = (-1)^{S_Q + S_q}$, where $S_q$ is the total spin of the light-quark pair that is created. Since for nonflip triplet decay models one has $S_q=1$, the selection rule becomes $CP = (-1)^{S_Q+1}$. According to this selection rule, decays into pairs of mesons with identical spatial wavefunctions are allowed for quarkonia, which have $CP=(-1)^{S_Q+1}$, and forbidden for quarkonium hybrids, which have $CP=(-1)^{S_Q}$.

Quarkonium hybrids in B\nobreakdash-O multiplets associated with bound states in the $\Pi_u$ potential, the $\Sigma_u^-$ potential, or coupled $\Pi_u$ and $\Sigma_u^-$ potentials have $CP=(-1)^{S_Q}$. The B\nobreakdash-O equivalent of two quark-model mesons with identical spatial wavefunctions is two heavy mesons in charge-conjugate HQSS doublets, e.g., two $S$-wave mesons, two $P^-$-wave mesons, or two $P^+$-wave mesons. Our finding that quarkonium hybrids in B\nobreakdash-O multiplets with a $\Sigma_u^-$ component may have nonzero decay rates into pairs of $S$-wave mesons therefore contradicts the selection rule $CP = (-1)^{S_Q+1}$, as first pointed out in Ref.~\cite{Bru23b}. This observation can be generalized to meson pairs other than $S+S$ by applying the selection rules in Section~\ref{selrulesec}. Bound states in the $\Sigma_u^-$ potential or coupled $\Pi_u$ and $\Sigma_u^-$ potentials can have nonzero decay rates into pairs of heavy mesons in charge-conjugate HQSS doublets with $j_1=j_2 \geq \frac{1}{2}$, such as $S+S$, $P^-+P^-$, or $P^++P^+$. Bound states in the $\Pi_u$ potential can have nonzero decay rates into pairs of heavy mesons in charge-conjugate HQSS doublets with $j_1=j_2 \geq \frac{3}{2}$, such as $P^++P^+$. These results support the assignment $S_q=0$ for quark-pair creation models of quarkonium-hybrid decays. In fact, contrarily to the widely used nonflip-triplet decay models with $S_q=1$, spin-singlet decay models with $S_q=0$ allow decays of quarkonium hybrids into meson pairs with identical spatial wavefunctions \cite{Bru19d}.

\subsection{Relative Partial Decay Rates}

The transition amplitudes in Eq.~\eqref{modelamp} depend on the spatial overlap of the quark-model wavefunctions only through the matrix elements $A$. The sums in Eq.~\eqref{modelamp} run over the two orbital-angular-momentum quantum numbers $L^\prime$ and $L^{\prime\prime}$. If the sums reduce to a single term, there is only one dominant spatial matrix element $A$ so it factors out of transition amplitudes of quarkonium states with the same quantum numbers $n$, $L_Q$, and all possible values of $S_Q$, $J$ into pairs of mesons with the same quantum numbers $n_1$, $n_2$, $L_1$, $L_2$, $L_Q^\prime$, and all possible values of $S_1$, $S_2$, $J_1$, $J_2$. In this case, nonflip triplet decay models can be used to predict the corresponding relative partial decay rates under the conditions outlined in Section~\ref{branratiosec}.

Conservation of parity and angular momentum, which require $(-1)^{L_Q + 1} = (-1)^{L_Q^\prime + L_1 + L_2}$ and triangle conditions for $(L^\prime, L_1, L_2)$, $(L^{\prime\prime},L^\prime,L_Q^\prime)$, and $(L^{\prime\prime}, L_Q, 1)$, respectively, imply that the sums over $L^\prime$ and $L^{\prime\prime}$ in Eq.~\eqref{modelamp} reduce to a single term if either:
\begin{enumerate}[a)]
\item $\lvert L_Q - L_Q^\prime\rvert = L_1 + L_2 + 1$,
\label{casea}
\item $\lvert L_1 - L_2 \rvert= L_Q + L_Q^\prime + 1$.
\label{caseb}
\end{enumerate}
Case \ref{casea}) corresponds to the difference $\lvert L_Q - L_Q^\prime\rvert$ being the maximum allowed by conservation of angular momentum for given values of $L_1$ and $L_2$. Case \ref{caseb}) corresponds to the difference $\lvert L_1 - L_2\rvert$ being the maximum allowed by conservation of angular momentum for given values of $L_Q$ and $L_Q^\prime$.

Outside of the two cases \ref{casea}) and \ref{caseb}), the transition amplitude $M$ in Eq.~\eqref{modelamp} is a linear combination of spatial transition matrix elements $A$. Then the relative partial decay rates are not completely determined by symmetries. This bears some resemblance to the situation in the B\nobreakdash-O approximation when there is more than one relevant radial transition amplitude $g_{\lambda,\eta}$.

It is possible to compare relative partial decay rates that are completely determined by symmetries both in nonflip triplet models and in the B\nobreakdash-O approximation using recoupling of angular momentum. The angular-momentum coefficients in Eq.~\eqref{modelcoef} are defined in terms of final-state mesons with total constituent-quark spins $S_1$ and $S_2$. The angular-momentum coefficients for decays into pairs of mesons with light-QCD angular momenta $j_1$ and $j_2$ can be defined by recoupling the angular momenta of the final-state mesons using Wigner 6-$j$ symbols:
\begin{multline}
\xi_{S,L_Q^\prime}^{L^\prime,L^{\prime\prime}}
\begin{bmatrix}
S_Q & L_Q & J \\
j_1 & L_1 & J_1 \\
j_2 & L_2 & J_2 \\
\end{bmatrix}_\pm
\equiv
(-1)^{L_1 + L_2 + J_1 + J_2} \sqrt{\tilde{\jmath}_1 \tilde{\jmath}_2} \\
\times
\sum_{S_1,S_2} \sqrt{\tilde{S}_1\tilde{S}_2}
\begin{Bmatrix}
\frac{1}{2} & \frac{1}{2} & S_1 \\
L_1 & J_1 & j_1 \\
\end{Bmatrix}
\begin{Bmatrix}
\frac{1}{2} & \frac{1}{2} & S_2 \\
L_2 & J_2 & j_2 \\
\end{Bmatrix}
\xi_{S,L_Q^\prime}^{L^\prime,L^{\prime\prime}}
\begin{bmatrix}
S_Q & L_Q & J \\
S_1 & L_1 & J_1 \\
S_2 & L_2 & J_2 \\
\end{bmatrix}_\pm.
\label{modelcoefmult}
\end{multline}
The square of the ratios of these angular-momentum coefficients give analytical approximations for the relative partial decay rates which can be directly compared with the B\nobreakdash-O predictions. The relative partial decay rates for $S$-wave quarkonia into $S+S$ ($P$-wave), $S+P^-$ ($S$-wave), and $S+P^+$ ($D$-wave) meson pairs in Table~\ref{quarkoniuml0} and those for $P$-wave quarkonia into $S+S$ ($S$-wave), $S+S$ ($D$-wave), and $S+P^+$ ($F$-wave) meson pairs in Table~\ref{quarkoniuml1} can also be predicted using nonflip triplet decay models. In these cases, it can be verified that nonflip triplet decay models and the B\nobreakdash-O approximation yield the same predictions for the same relative partial decay rates.

In general instances of cases \ref{casea}) and \ref{caseb}), the angular-momentum coefficients of nonflip triplet decay models obtained by inserting Eq.~\eqref{modelcoef} into Eq.~\eqref{modelcoefmult} and those of the B\nobreakdash-O approxiation obtained by inserting Eq.~\eqref{mixpot1} into Eq.~\eqref{mixpot2} differ by multiplicative factors. The squares of these factors, however, do not depend on the quantum numbers $S_Q$ and $J$ of the initial quarkonium state nor the quantum numbers $J_1$, $J_2$, and $S$ of the final meson-pair state. Therefore, they factor out of relative partial decay rates that are completely determined by symmetries. The angular-momentum coefficients of nonflip triplet decay models actually agree (up to signs) with those of the B\nobreakdash-O approximation only in simple instances involving two $S$-wave mesons in the final state or, alternatively, one $S$-wave meson in the final state and either $L_Q=0$ or $L_Q^\prime=0$.

Note that the B\nobreakdash-O approximation is generally more predictive than nonflip triplet decay models because it incorporates cylindrical symmetries. That is exemplified by the decays of $P$-wave quarkonia into one $S$-wave meson and one $P$-wave meson in a relative $P$-wave. In nonflip triplet decay models, the relative partial decay rates are not completely determined by symmetries since these decays do not fall into either case \ref{casea}) or \ref{caseb}). The B\nobreakdash-O symmetries, on the other hand, fully constrain the relative partial decay rates for $P$-wave quarkonia into $S+P^-$ ($P$-wave) and $S+P^+$ ($P$-wave) to the values listed in Table~\ref{quarkoniuml1}.

While the B\nobreakdash-O approximation reproduces relative partial decay rates from nonflip-triplet models of quarkonium decays, we find no support for such models of quarkonium hybrid decays. For quarkonium hybrids, the B\nobreakdash-O angular-momentum coefficients and relative partial decay rates differ from those of nonflip-triplet decay models even in simple cases. This disagreement is in addition to the already mentioned violation of the quark-pair-creation-model selection rule that prohibits decays of quarkonium hybrids into pairs of mesons with identical spatial wavefunctions.

\section{Overview}
\label{conclusionsec}    

In this work, we have derived model-independent results for the decays of hidden-heavy hadrons into pairs of heavy hadrons using the diabatic B\nobreakdash-O approximation for QCD. Our general expressions for the coupling potentials between a hidden-heavy hadron and heavy-hadron pairs in Eq.~\eqref{mixpot1} are sums of products of B\nobreakdash-O transition amplitudes and angular-momentum coefficients. The B\nobreakdash-O transition amplitudes $g_{\lambda,\eta}$ can be calculated using lattice QCD with static $\bm{3}$ and $\bm{3}^\ast$ color sources.  The angular-momentum coefficients in Eq.~\eqref{mixpot2} are determined by products of Clebsch-Gordan coefficients, Wigner 6-$j$ symbols, and Wigner 9-$j$ symbols.

We have used these expressions to derive selection rules for the decays of hidden-heavy hadrons into pairs of heavy hadrons. The selection rules are given in Section~\ref{selrulesec}. We also obtained analytic approximations for relative partial decay rates between hidden-heavy hadrons in the same B\nobreakdash-O multiplet and pairs of heavy hadrons in specific heavy-quark-spin-symmetry doublets. The relative partial decay rates are given in Eq.~\eqref{partdecays}. Our results reproduce and extend previous ones obtained for quarkonium decays using heavy-quark spin symmetry alone in Refs.~\cite{Vol13,DeR77,Kai03}. Model-independent results have been obtained previously for exotic hidden-heavy mesons under the assumption that they are heavy-meson molecules \cite{Vol04,Bon11,Vol12,Vol16,Liu13}. Our methods can be used to extend those results.

We have discussed in detail the selection rules and relative partial decay rates for decays of quarkonia and quarkonium hybrids into $S+S$, $S+P^-$, and $S+P^+$ meson pairs. For decays of $S$-wave quarkonia into $S+S$, $S+P^-$, and $S+P^+$ meson pairs and for decays of $P$-wave quarkonia into $S+S$ meson pairs and into $S+P^+$ meson pairs in a relative $F$-wave, our results reproduce previous results obtained using quark-pair-creation models \cite{LeY88,Bur14}. For decays of $P$-wave quarkonia into $S+P^-$ meson pairs in a relative $P$-wave and into $S+P^+$ meson pairs in a relative $P$-wave and for decays of quarkonium hybrids into $S+S$, $S+P^-$, and $S+P^+$ meson pairs, our results differ from those of quark-pair-creation models. The prediction of nonvanishing partial decay rates for some quarkonium hybrids into $S+S$ meson pairs, which was first pointed out in Ref.~\cite{Bru23b}, contradicts the conventional wisdom for the last 40 years from quark-pair-creation models \cite{Tan82,LeY84,Is85,Idd88,Ish91,Clo94,Pag97,Pag99,Kou05}. There is some evidence from lattice QCD that the conventional selection rule is violated by decays of the $1^{-+}$ charmonium hybrid state \cite{Shi23}.

It is important to emphasize that the results of this paper do not apply exclusively to the decays of quarkonia and quarkonium hybrids into pairs of heavy mesons. They can be equally well applied to their decays into a pair of heavy baryons. They can also be applied to the decays of other exotic hidden-heavy hadrons, including hidden-heavy tetraquark mesons and hidden-heavy pentaquark baryons. All it takes is to identify the B\nobreakdash-O quantum numbers of the decaying hidden-heavy hadron and to specify the light-QCD quantum numbers of the two heavy hadrons in the final state. One can derive the corresponding selection rules by applying our general expressions in Section~\ref{selrulesec}. If the decays are dominated by a single B\nobreakdash-O transition amplitude, the ratio of partial decay rates is given by our general analytic expression in Eq.~\eqref{partdecays}.

Similar methods can be applied to the decays of double-heavy hadrons into pairs of heavy hadrons. A double-heavy hadron contains two heavy quarks with the same flavor. The B\nobreakdash-O transition amplitude for its  decays into pairs of heavy hadrons can be calculated using lattice QCD with two static $\bm{3}$ color sources. The selection rules and relative partial decay rates for decays of double-heavy hadrons are more intricate than those for decays of hidden-heavy hadrons, because light flavor symmetries play a more significant role \cite{inprep}.

The expressions for the coupling potentials obtained by inserting Eq.~\eqref{mixpot1} into Eq.~\eqref{mixpot2} can be combined with the techniques pioneered in Ref.~\cite{Bru23a} to include spin splittings of heavy hadrons in the diabatic B\nobreakdash-O framework for hidden-heavy hadrons coupled with heavy-hadron pairs. That is extremely important for exotic hidden-heavy hadrons, since being able to resolve heavy-hadron spin splittings grants access to the near-threshold dynamics that characterizes many exotic states. Using the techniques of Ref.~\cite{Bru23a} together with those of this paper, calculating the spectrum of both conventional and exotic hidden-heavy hadrons boils down to calculating just a handful of B\nobreakdash-O potentials and transition amplitudes using lattice QCD with static $\bm{3}$ and $\bm{3}^\ast$ color sources.  We strongly encourage theoretical efforts in this direction.

\acknowledgments{%
This research was supported by the U.S. Department of Energy under Grant No.\ DE-SC0011726. This work contributes to the goals of the US DOE ExoHad Topical Collaboration, Contract No.\ DE-SC0023598.%
}

\bibliography{decaysbib}

\end{document}